\documentclass[twoside,british,english,footinbib,prb,twocolumn,superscriptaddress]{revtex4}
\usepackage[T1]{fontenc}
\usepackage[latin1]{inputenc}
\usepackage{babel}
\usepackage{graphics}

\makeatletter

\providecommand{\LyX}{L\kern-.1667em\lower.25em\hbox{Y}\kern-.125emX\@}

\makeatother
\begin{document}

\title{Exotic pairing \textit{via} a central attraction: from the BCS to
the Bose limits}

\author{Jorge Quintanilla}

\email{quintanilla@if.sc.usp.br}

\homepage{http://www.if.sc.usp.br/~quintanilla}

\affiliation{H.H. Wills Physics Laboratory, University of Bristol, Tyndall Av.,
Bristol BS8 1TL, U.K.}

\affiliation{Departamento de F\'{\i}sica en Inform\'atica, Instituto de F\'{\i}sica
de S\~{a}o Carlos, Universidade de S\~{a}o Paulo, Caixa Postal 369,
S\~{a}o Carlos SP 13560-970, Brazil}

\author{Balazs L. Gy\"orffy}

\affiliation{H.H. Wills Physics Laboratory, University of Bristol, Tyndall Av.,
Bristol BS8 1TL, U.K.}

\author{James F. Annett}

\affiliation{H.H. Wills Physics Laboratory, University of Bristol, Tyndall Av.,
Bristol BS8 1TL, U.K.}

\author{Jonathan P. Wallington}

\affiliation{H.H. Wills Physics Laboratory, University of Bristol, Tyndall Av.,
Bristol BS8 1TL, U.K.}

\begin{abstract}
In the context of a simple model featuring an explicit, \emph{central}
interaction potential, and using a standard functional-integral technique,
we study superconductivity with angular momentum quantum number \( l=2 \)
as an emergent property of the many-body system. Our interaction potential
is attractive at a finite distance \( r_{0} \), and the breaking
of the rotational symmetry is the result of an interplay between \( r_{0} \)
and the inter-particle distance \( r_{s} \) which we deem generic
to interactions of this type. However such interplay, responsible
for the preference of a \( d \)-wave state for a range of intermediate
densities, takes place only in the BCS limit. In contrast, as the
Bose-Einstein (BE) limit is approached the internal energy of the
{}``preformed pairs'' becomes the dominant contribution and there
is a quantum phase transition in which the \( s \)-wave symmetry
is restored. We also find that the limiting value of the critical
temperature is \( k_{B}T_{c}\rightarrow 3.315\, \hbar ^{2}/2m^{*}\, \left[ n/2\left( 2l+1\right) \right] ^{2/3}, \)
which coincides with the usual result only for \( l=0 \); for \( l>0 \),
it differs in the degeneracy factor \( 1/\left( 2l+1\right)  \),
which lowers \( T_{c} \). Our results thus place constraints on exotic
pairing in the BE limit, while at the same time indicating a particularly
interesting route to pairing with \( l>0 \) in a BCS superconductor. 
\end{abstract}
\maketitle

\section{Introduction}

It is a surprising consequence of BCS theory \cite{Bardeen-Cooper-Schrieffer-57}
that certain \emph{central} interaction potentials \( V\left( \left| \mathbf{r}-\mathbf{r}'\right| \right)  \)
lead to Cooper pairing with a finite value of the angular momentum,
thus breaking the rotational symmetry of the continuum \cite{Balian-64}.
The phenomenon is analogous to what happens in {}``Hubbard'' models
with attraction between nearest-neighbours for which, as is well known
\cite{Micnas-Ranninger-Robaskiewicz-90}, \( d \)-wave pairing can
break the symmetry of the crystal lattice. However, it is especially
interesting in the original context \cite{Balian-64} of a continuum
model, because of the contrast with the well-known theorem \cite{Landau-Lifshitz-QM}
for two-body pairing in real space, which demands that the ground
state minimizes the orbital angular momentum. This rotational symmetry
breaking is thus a many-body effect, and one expects that in the limit
of low densities and strong attraction, when the BCS ground state
is a Bose-Einstein (BE) condensate of non-overlapping pairs \cite{Eagles-69,Pincus-et-al-73,Leggett-80,Nozieres-SchmittRink-85},
the rotational symmetry of the system is restored. However until now
investigations of the BCS to Bose crossover for non \( s \)-wave
pairing were performed either in the context of lattice models (see
Ref.~\onlinecite{Micnas-Ranninger-Robaskiewicz-90} for a review and
Refs.~\onlinecite{Engelbrecht-Nazarenko-Randeria-98,DenHertog-99,Andrenacci-Perali-Pieri-Strinati-99,Jon-James-2000,Chen-Kosztin-Levin-00,Soares-Kokubun-RodriguezNunez-Rendon-02}
for some examples of recent work) or for the anisotropic interaction
potential of Ref.~\onlinecite{Zwerger-97}.%
\footnote{For fermions in a continuum with a repulsive interaction \( V\left( r\right)  \),
the Kohn-Luttinger mechanism \cite{Kohn-Luttinger-65} can lead to
pairing with high angular momentum \cite{Chubukov-Kagan-89}. But
note that in this case the BE limit can never be realised, since obviously
there is no two-body bound state.
} 

In this paper we take a slightly different approach by studying a
continuum model, but choosing to work with an explicit, \emph{central}
interaction potential \( V\left( \left| \mathbf{r}-\mathbf{r}'\right| \right)  \)
which can lead to pairing with more than one value of the angular
momentum: the {}``delta-shell'' potential \cite{Gottfried-66,Villarroel-98}.
The resulting {}``delta-shell'' model (DSM) captures, in an idealised
way, the essential feature leading to Cooper pairing with finite value
of the angular momentum, namely being attractive at a well-defined,
finite distance \cite{Quintanilla-Gyorffy-00}. Thus we expect some
of the novel features that we shall describe, pertaining to the mechanisms
by which the rotational symmetry is broken in the BCS limit and restored
in the BE limit, to be generic to a large class of central effective
interactions. In particular we shall see that for such models the
evolution of a BCS superconductor with exotic pairing towards the
BE limit involves a phase transition in which the symmetry of the
superconducting order parameter is increased. This adds to the work
by Babaev and Kleinert \cite{Babaev-Kleinert-98} who also found,
in the context of a chiral Gross-Neveu model, a phase transition associated
with the BCS to Bose crossover. However the nature of the phase transition
that we describe here is quite different, as it takes place in the
superconducting state, while that of Babev and Kleinert corresponds
to the formation of preformed pairs in the normal state.

\section{The Delta-Shell Model}

The first discussions of exotic Cooper pairing \cite{Balian-64} took
place in the context of the weak-coupling theory of superfluid \( ^{3}\textrm{He} \).
It was assumed that there existed a central, non-retarded interaction
potential \( V\left( \left| \mathbf{r}-\mathbf{r}'\right| \right)  \)
acting between particles at positions \( \mathbf{r} \) and \( \mathbf{r}' \).
One then writes \begin{equation}
\label{angle decomp of V}
V\left( \mathbf{k}-\mathbf{k}'\right) =\sum _{l'=0}^{\infty }K_{l'}\left( \left| \mathbf{k}\right| ,\left| \mathbf{k}'\right| \right) \left( 2l'+1\right) P_{l'}\left( \hat{\mathbf{k}}.\hat{\mathbf{k}}'\right) ,
\end{equation}
 \foreignlanguage{british}{}where \( V\left( \mathbf{k}-\mathbf{k}'\right) \equiv \int d^{3}\mathbf{r}\, e^{i\left( \mathbf{k}-\mathbf{k}'\right) .\mathbf{r}}\, V\left( \mathbf{r}-\mathbf{r}'\right)  \),
and finds that each of the terms in this series leads to pairing with
a different value of the angular momentum quantum number, \( l \).
As it can, and has been, argued, in the weak-coupling limit one can
approximate\begin{equation}
\label{angle decomp of V with wc approx}
V\left( \mathbf{k}-\mathbf{k}'\right) \approx K_{l}\, \left( 2l+1\right) P_{l}\left( \hat{\mathbf{k}}.\hat{\mathbf{k}}'\right) 
\end{equation}
where \( l \) is the value of \( l' \) for which the coupling constant
on the Fermi surface, \begin{equation}
\label{BCS limit coupling const}
K_{l'}\equiv K_{l'}\left( k_{F},k_{F}\right) ,
\end{equation}
 is largest. The approximate form (\ref{angle decomp of V with wc approx})
of the potential \( V\left( \mathbf{k}-\mathbf{k}'\right)  \) is,
for \( l>0 \), anisotropic, and it leads to pairing with finite angular
momentum quantum number \( l \) (see Ref.~\onlinecite{Balian-64}).
For \( l=0 \), it reduces to the BCS {}``contact potential'' \cite{Bardeen-Cooper-Schrieffer-57},
leading to \( s \)-wave pairing. Although introduced in the context
of a weak-coupling theory, the contact potential has often been used
to study the BCS to Bose crossover \cite{Randeria-Duan-Shieh-90,Zwerger-92,SaDeMelo-Randeria-Engelbrecht-93,Engelbrecht-Randeria-SaDeMelo-97,Marini-Pistolesi-Strinati-98,Haussmann-93,Haussmann-94,Babaev-Kleinert-99,Pistolesi-Strinati-94,Pistolesi-Strinati-96}.
Similarly, Stintzing and Zwerger have considered a simplified potential
of the form (\ref{angle decomp of V with wc approx}) with \( l=2 \)
to study the BCS to Bose crossover for pairs with \( d_{x^{2}-y^{2}} \)
symmetry \cite{Zwerger-97} (but in two dimensions, and with the additional
assumption of separability to make it more tractable). One of the
key results of this later work \cite{Zwerger-97} was that the critical
temperature is given, in the BE limit, by the same expression as in
the \( s \)-wave case \cite{Nozieres-SchmittRink-85,Zwerger-92,SaDeMelo-Randeria-Engelbrecht-93}:
\textit{\begin{equation}
\label{TcBE Ogg}
k_{B}T_{c}\approx 3.315\frac{\hbar ^{2}}{2m^{*}}\left( \frac{n}{2}\right) ^{2/3}\, {\rm \, for}\, s\, {\rm \, and}\, d_{x^{2}-y^{2}}\, {\rm \, pairing}
\end{equation}
}

Although very useful, the above approach is not appropriate to study
the question that we are interested in here, since it introduces a
particular pairing symmetry at the level of the interaction potential.
In contradistinction, we want to find pairing with \( l>0 \) as an
emergent property of the many-body system. Moreover, we would expect,
on the basis of the above arguments, to recover \( l=0 \) pairing
in the BE limit, in which the internal structures of the Cooper pairs
are independent. This physics seems also to be absent from those studies,
as Eq.~(\ref{TcBE Ogg}) suggests that the critical temperature is
degenerate for \( s \) and \( d_{x^{2}-y^{2}} \) superconductivity. 

An alternative strategy is to do the calculations taking the full
\( r \)-dependence of \( V\left( r\right)  \) into account. A study
of this type was carried out by Andrenacci \textit{et al.} \cite{Andrenacci-Perali-Pieri-Strinati-99}
who took a Gaussian form for \( V\left( r\right)  \). This allowed
them to investigate the properties of the crossover at finite densities
(in contrast, as is well known, the procedure required to regulate
the ultraviolet divergences associated with simplified potentials
of the form (\ref{angle decomp of V with wc approx}) at all couplings
is only valid in the dilute limit \cite{Randeria-95}). They also
considered the highly idealised separable potential introduced in
the seminal paper by Nozi\`eres and Schmitt Rink \cite{Nozieres-SchmittRink-85},
which has been employed in several other instances \cite{Pistolesi-Strinati-94,Pistolesi-Strinati-96}
on account of its mathematical simplicity (but note that this is not,
strictly speaking, a central potential). However in either case there
was no rotational symmetry breaking: even at high densities they only
obtained \( s \)-wave superconductivity (the discussion of \( d_{x^{2}-y^{2}} \)
superconductivity in Ref.~\onlinecite{Andrenacci-Perali-Pieri-Strinati-99}
is based on a lattice model). 

On the other hand, a simple argument \cite{Quintanilla-Gyorffy-00}
based on the BCS {}``gap equation'' suggests that \( l>0 \) Cooper
pairing is associated with central potentials \( V\left( r\right)  \)
that are non-monotonic functions of \( r \), with maximum attraction
near some finite distance, \( r\sim r_{0}>0 \). The delta-shell potential
was proposed in Ref.~\onlinecite{Quintanilla-Gyorffy-00} as the simplest
form of \( V\left( r\right)  \) that has this feature:\begin{equation}
\label{delta-shell potential}
V\left( \left| \mathbf{r}-\mathbf{r}'\right| \right) =-g\delta \left( \left| \mathbf{r}-\mathbf{r}'\right| -r_{0}\right) 
\end{equation}
The resulting DSM can be regarded as the continuum analogue of the
lattice model with nearest-neighbour attraction discussed in Refs.~\onlinecite{Micnas-Ranninger-Robaskiewicz-90,DenHertog-99,Andrenacci-Perali-Pieri-Strinati-99,Jon-James-2000,Soares-Kokubun-RodriguezNunez-Rendon-02},
for example. But note that in the DSM the distance \( r_{0} \) at
which the fermions attract each other is a free parameter that can
be varied continuously, and the non-interacting dispersion relation
is that of free fermions with an effective mass \( m^{*} \).

\begin{figure}
{\centering \resizebox*{0.9\columnwidth}{!}{\includegraphics{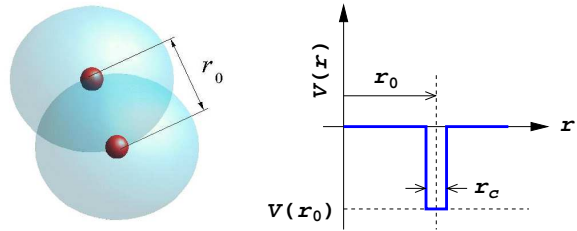}} \par}

\caption{\label{Fig1}The delta-shell interaction potential. \emph{Left}\textbf{:}
the two particles attract each other only when each of them lay on
a thin shell, of radius \protect\( r_{0}\protect \), centered on
the other one. \emph{Right:} The delta-shell interaction potential
can be regarded as an approximation to any central potential that
is attractive only near some distance \protect\( r_{0}\protect \)
(see text).}
\end{figure}

The delta-shell potential can also be considered an approximation
to any central potential that is attractive only within a range of
distances centred at \( r_{0} \), of width \( r_{c}\ll r_{0} \),
since Eq.~(\ref{delta-shell potential}) is equivalent to performing,
in the general expression \begin{equation}
\label{kernel in general}
K_{l}\left( \left| \mathbf{k}\right| ,\left| \mathbf{k}'\right| \right) =\left( -1\right) ^{l}\int _{0}^{\infty }dr4\pi r^{2}j_{l}\left( \left| \mathbf{k}\right| r\right) V\left( r\right) j_{l}\left( \left| \mathbf{k}'\right| r\right) ,
\end{equation}
valid for any central potential (\( j_{l}\left( x\right)  \) denotes
a spherical Bessel function), the approximation\begin{eqnarray}
\int _{0}^{\infty }dr\, 4\pi r^{2}\, j_{l}\left( \left| \mathbf{k}\right| r\right) \, V\left( r\right) \, j_{l}\left( \left| \mathbf{k}'\right| r\right)  &  & \nonumber \\
 &  & \hspace {-4.5cm}\approx r_{c}\, 4\pi r_{0}^{2}\, j_{l}\left( \left| \mathbf{k}\right| r_{0}\right) \, V\left( r_{0}\right) \, j_{l}\left( \left| \mathbf{k}'\right| r_{0}\right) \label{ds kernels} 
\end{eqnarray}
which corresponds to taking the limit \( r_{c}\rightarrow 0 \) while
keeping \( V\left( r_{0}\right) \times r_{c}=constant\equiv -g \)
(\( g \) thus has dimensions of \( energy\times length \)). A particularly
simple example of this is the square well of Fig.~\ref{Fig1}.

The two-body problem associated with the delta-shell potential is
very well-known (see Refs.~\onlinecite{Gottfried-66,Villarroel-98},
for example). In particular, it can bind a pair in free space with
any value of \( l=0,1,2,\ldots  \) To simplify matters, we will assume
that the attraction takes place between particles with opposite spins.
Finally, in \( \mathbf{k} \)-space the delta-shell potential is given
by \( V\left( \mathbf{k}-\mathbf{k}'\right) =-g4\pi r_{0}^{2}\sin \left( \left| \mathbf{k}-\mathbf{k}'\right| r_{0}\right) /\left| \mathbf{k}-\mathbf{k}'\right| r_{0} \),
from which it is evident that it reduces to the contact potential
in the limit \( r_{0}\rightarrow 0 \) (keeping \( g4\pi r_{0}^{2} \)
equal to \( -K_{0} \)). Interestingly, the delta-shell potential,
for any finite \( r_{0} \), does not display the ultraviolet divergences
affecting the contact potential. 

We will study the BCS to Bose crossover in this new model using the
standard functional-integral technique of Refs.~\onlinecite{Zwerger-92,Alexandrov-Rubin-93,SaDeMelo-Randeria-Engelbrecht-93,Randeria-95}.
At zero temperature, it implies a description of the system in the
saddle-point approximation, which amounts to using the BCS ground
state \cite{Randeria-95} (as in Refs.~\onlinecite{Leggett-80,Nozieres-SchmittRink-85,DenHertog-99,Andrenacci-Perali-Pieri-Strinati-99,Soares-Kokubun-RodriguezNunez-Rendon-02},
for example). Thus our results for the ground state will be approximate,
but of variational significance. At the critical temperature, Gaussian
fluctutations about the saddle point are taken into account as in
Refs.~\onlinecite{Zwerger-92,SaDeMelo-Randeria-Engelbrecht-93,Zwerger-97}
(as is well-known \cite{Randeria-95} this is equivalent to the RPA-like
diagrammatic technique introduced by Nozi\`eres and Schmitt-Rink \cite{Nozieres-SchmittRink-85}).
This approach is rather limited in that, in the strong-coupling limit,
it neglects interactions between the preformed pairs, existing above
\( T_{c} \), and so it can only describe the effect of fluctuations
on the superconducting instability at low densities \cite{Haussmann-93,Haussmann-94}.
Nevertheless as we shall see it is enough to discuss the rotational
symmetry breaking in the weak-coupling limit, in which the fluctuations
are negligible, as well as the mechanism by which the critical temperature
becomes larger for \( s \)-wave pairs in the BE limit (at strong
coupling \emph{and} low densities). The application of these standard
methods to the DSM is fairly straight-forward, so we will quote here
only the key expressions; %we have outlined the derivation for the DSM in an appendix, and further
details can be found in Ref.~\onlinecite{Quintanilla-01}.

Our model has four parameters: the distance \( r_{0} \) at which
the attraction takes place, the coupling constant \( g \), the density
of fermions \( n \) and their mass \( m^{*} \). In principle, the
BCS ground state can be characterized by the dependence of the zero-temperature
gap function \( \Delta _{\mathbf{k}} \) and chemical potential \( \mu  \)
on these four parameters. Likewise, the superconducting instability
can be described by giving the critical values of the temperature
\( T_{c} \) and chemical potential \( \mu _{c} \) in terms of \( r_{0},g,n,m^{*} \).
However the DSM has the remarkable property that the four energies
\( \Delta _{\mathbf{k}},\, \mu ,\, k_{B}T_{c}\, ,\mu _{c} \), rescaled
by the {}``localisation energy'' \( \varepsilon _{0}\equiv \hbar ^{2}/2m^{*}r_{0}^{2} \)
(which we will denote \( \tilde{\Delta }_{\tilde{\mathbf{k}}},\, \tilde{\mu },\, \tilde{T}_{c}\, ,\tilde{\mu }_{c} \),
where \( \tilde{\mathbf{k}}\equiv r_{0}^{-1}\mathbf{k} \)), are functions
of only two parameters: the {}``dimensionless coupling constant''
\( \tilde{g}\equiv \left( \varepsilon _{0}r_{0}\right) ^{-1}\times g \)
and the number of fermions in a sphere of radius \( r_{0} \): \( \tilde{n}\equiv \left( 4\pi /3\right) r_{0}^{3}\times n \).

\section{\label{SEC-ground state}Ground state}

The kernel \( K_{l'}\left( \left| \mathbf{k}\right| ,\left| \mathbf{k}'\right| \right)  \)
in Eq.~(\ref{angle decomp of V}) is given by \begin{equation}
\label{V separable DSM}
K_{l}\left( \left| \mathbf{k}\right| ,\left| \mathbf{k}'\right| \right) =-g4\pi r_{0}^{2}\left( -1\right) ^{l}j_{l}\left( \left| \mathbf{k}\right| \right) j_{l}\left( \left| \mathbf{k}'\right| \right) 
\end{equation}
Thus although evidently the delta-shell potential is central and therefore
not separable in the sense of the NSR potential, it can be written
as a sum of separable terms, each one corresponding to a different
value of \( l \). Accordingly the gap function \( \tilde{\Delta }_{\tilde{\mathbf{k}}} \)
, at the saddle point, has the following form:\[
\tilde{\Delta }_{\tilde{\mathbf{k}}}\equiv \sum _{l=0}^{\infty }\sum _{m=-l}^{l}\tilde{\Delta }_{l,m}j_{l}\left( \left| \mathbf{k}\right| r_{0}\right) Y_{l,m}\left( \hat{\mathbf{k}}\right) \]
 In terms of the amplitudes \( \tilde{\Delta }_{l,m} \) the usual
{}``gap'' and {}``density'' equations, for a homogeneous, stationary,
non-polarised state with singlet pairing (\( \tilde{\Delta }_{l,m}\equiv 0 \)
for odd \( l \)), read\begin{eqnarray}
\tilde{\Delta }_{lm} & = & \sum _{l',m'}\left\{ \int \frac{d^{3}\tilde{\mathbf{k}}}{\left( 2\pi \right) ^{3}}\frac{\tilde{g}\tilde{\Lambda }_{lm,l'm'}\left( \tilde{\mathbf{k}}\right) }{2\tilde{E}_{\tilde{\mathbf{k}}}}\right\} \tilde{\Delta }_{l'm'}\label{rescaled-gap-eq} \\
\tilde{n} & = & \int \frac{d^{3}\tilde{\mathbf{k}}}{6\pi ^{2}}\left( 1-\frac{\tilde{\varepsilon }_{\tilde{\mathbf{k}}}}{\tilde{E}_{\tilde{\mathbf{k}}}}\right) \label{rescaled-n-eq} 
\end{eqnarray}
where \( \tilde{\varepsilon }_{\tilde{\mathbf{k}}}\equiv \tilde{\mathbf{k}}^{2}-\tilde{\mu } \),
\( \tilde{E}_{\tilde{\mathbf{k}}}\equiv \sqrt{\tilde{\varepsilon }_{\tilde{\mathbf{k}}}+\left| \tilde{\Delta }_{\tilde{\mathbf{k}}}\right| ^{2}} \)
and \[
\tilde{\Lambda }_{lm,l'm'}\left( \tilde{\mathbf{k}}\right) \equiv \left( 4\pi \right) ^{2}j_{l}\left( \left| \tilde{\mathbf{k}}\right| \right) j_{l'}\left( \left| \tilde{\mathbf{k}}\right| \right) Y^{*}_{lm}\left( \hat{\mathbf{k}}\right) Y_{l'm'}\left( \hat{\mathbf{k}}\right) \]
To assess the relative stability of different solutions to the self-consistency
problem (\ref{rescaled-gap-eq},\ref{rescaled-n-eq}), corresponding
to the same values of \( \tilde{g} \) and \( \tilde{n} \), one has
to compare the corresponding ground-state energies: \begin{equation}
\label{rescaled-internal-energy-equation}
\tilde{\mathcal{U}}=\int \frac{d^{3}\tilde{\mathbf{k}}}{6\pi ^{2}}\left| \tilde{\mathbf{k}}\right| ^{2}\left( 1-\frac{\tilde{\varepsilon }_{\tilde{\mathbf{k}}}}{\tilde{E}_{\tilde{\mathbf{k}}}}\right) -\sum _{l,m}\frac{\left| \tilde{\Delta }_{l,m}\right| ^{2}}{12\pi \tilde{g}}-\frac{3}{4}\tilde{g}\tilde{n}^{2}
\end{equation}

\begin{figure}
{\centering \resizebox*{0.9\columnwidth}{!}{\includegraphics{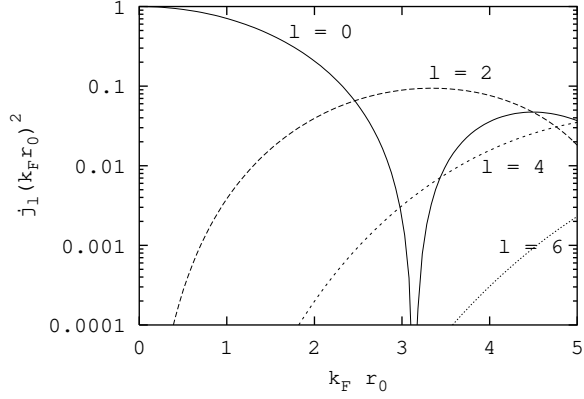}} \par}

\caption{\label{FIG-weighting factor}The strength of the attraction in the
BCS limit for pairing with the first four even values of the angular
momentum quantum number, \protect\( l=0,2,4,6\protect \).}
\end{figure}

Evidently Eq.~(\ref{rescaled-gap-eq}) is an infinite system of non-linear
integral equations with, presumably, an infinite number of non-trivial
solutions and there is no systematic way of finding all of them. Nevertheless
a certain subset, selected by the requirement that all but a few of
the \( \tilde{\Delta }_{l,m} \) are equal to zero, can be explored
systematically. Since the effective coupling constant in the weak-coupling
limit, Eq.~(\ref{BCS limit coupling const}), is (for even \( l \))
\begin{equation}
\label{BCS limit coupling const DSM}
K_{l}=-g4\pi r_{0}^{2}\, j_{l}\left( k_{F}r_{0}\right) ^{2}
\end{equation}
it is clear that for small \( \tilde{g} \), and within the range
of densities for which \( k_{F}r_{0}\alt 5 \), we can restrict our
attention to the first two values of \( l=0,2 \) (see Fig.~\ref{FIG-weighting factor}).
These are the two-body states with lowest energy (existing at \( \tilde{g}\ge 2,10 \),
respectively \cite{Gottfried-66,Villarroel-98}), and therefore this
simplification is also valid for our purposes in the BE limit. Moreover,
for simplicity we will consider \( d \)-wave states with a particular
symmetry, choosing \( d_{x^{2}-y^{2}} \) which has been extensively
studied in other models \cite{Micnas-Ranninger-Robaskiewicz-90,Zwerger-97,DenHertog-99,Andrenacci-Perali-Pieri-Strinati-99,Soares-Kokubun-RodriguezNunez-Rendon-02,Jon-James-2000}
on account of its relevance to cuprate superconductivity \cite{Annett-Goldenfeld-Leggett-96}.

\begin{figure*}
{\centering \resizebox*{!}{0.5\textheight}{\includegraphics{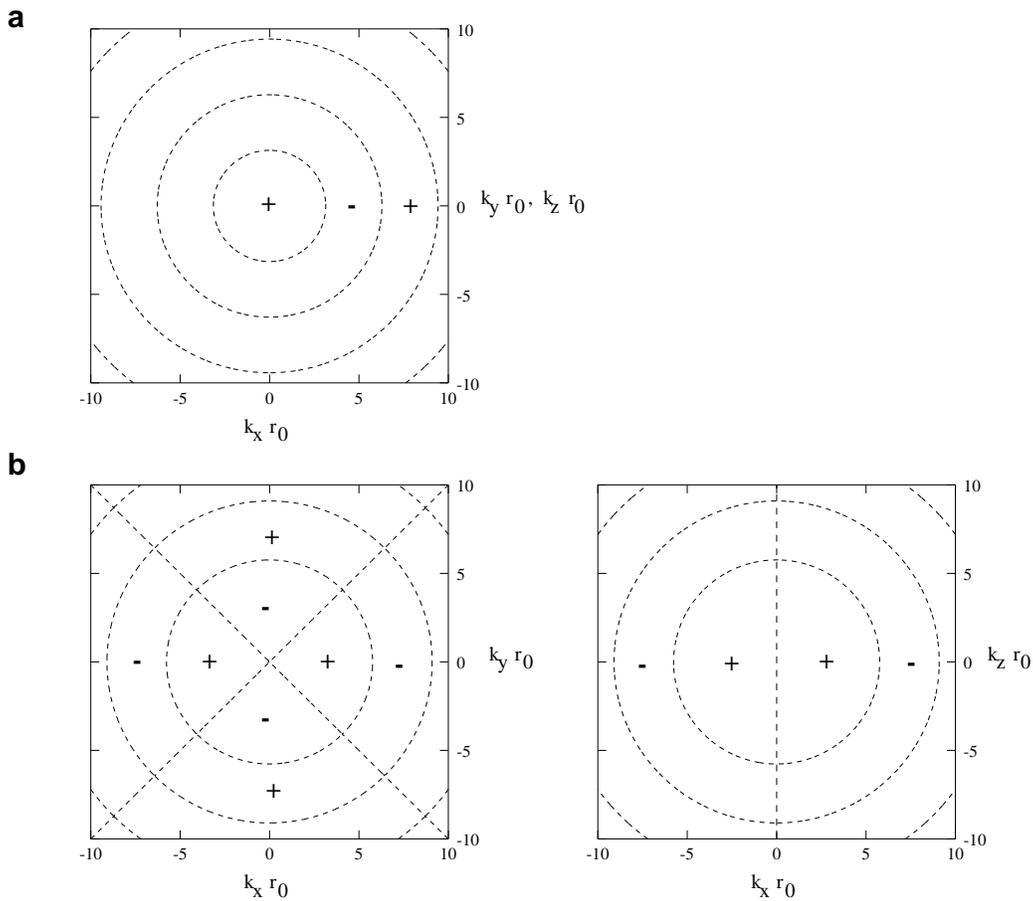}} \par}

\caption{\label{FIG-nodal structure}The zeroes (dashed lines) and sign ({}``+''
and {}``-'' symbols) of the gap function \protect\( \Delta _{\mathbf{k}}\protect \)
on the \protect\( (k_{x},k_{y})\protect \) and \protect\( (k_{x},k_{z})\protect \)
planes, for (a) the trial ground state with \protect\( s\protect \)
symmetry and (b) the one with \protect\( d_{x^{2}-y^{2}}\protect \)
symmetry.}
\end{figure*}

For our two trial ground states the gap function has the following
form, respectively:\begin{eqnarray}
\tilde{\Delta }_{s}\left( \tilde{\mathbf{k}}\right)  & \equiv  & \tilde{\Delta }_{s}j_{0}\left( \left| \tilde{\mathbf{k}}\right| \right) Y_{00}\label{s-wave-gf} \\
\tilde{\Delta }_{d_{x^{2}-y^{2}}}\left( \tilde{\mathbf{k}}\right)  & \equiv  & \tilde{\Delta }_{d_{x^{2}-y^{2}}}j_{2}\left( \left| \tilde{\mathbf{k}}\right| \right) \times \label{dxmy2-wave-gf} \\
 &  & \times \frac{1}{\sqrt{2}}\left[ Y_{2,2}\left( \hat{\mathbf{k}}\right) +Y_{2,-2}\left( \hat{\mathbf{k}}\right) \right] \nonumber 
\end{eqnarray}
Note that this is a more complicated \( \mathbf{k} \)-dependence
than that of gap functions arising from interactions of the form (\ref{angle decomp of V with wc approx}),
which depend only on the angle \( \hat{\mathbf{k}} \). In particular,
the gap function can change sign as \( \mathbf{k} \) increases in
the \emph{radial} direction of increasing \( \left| \mathbf{k}\right|  \),
not just as the angle \( \hat{\mathbf{k}} \) is varied: see Fig.~\ref{FIG-nodal structure}.
The oscillatory behaviour as a function of \( \left| \mathbf{k}\right|  \)
can be regarded as a direct consequence of the singling out of a particular
distance by the attractive interaction (\ref{delta-shell potential}).
More generally, we expect these oscillations, of frequency \( \sim 1/r_{0} \),
to be a generic feature of interactions that are attractive predominantly
at some finite distance \( r_{0} \).

Substitution of Eq.~(\ref{s-wave-gf}) (Eq.~(\ref{dxmy2-wave-gf}))
into the self-consistency problem (\ref{rescaled-gap-eq},\ref{rescaled-n-eq})
yields a much simpler problem, which can be solved numerically for
every value of \( \tilde{g} \) and \( \tilde{n} \).

\begin{figure}
{\centering \resizebox*{0.9\columnwidth}{!}{\includegraphics{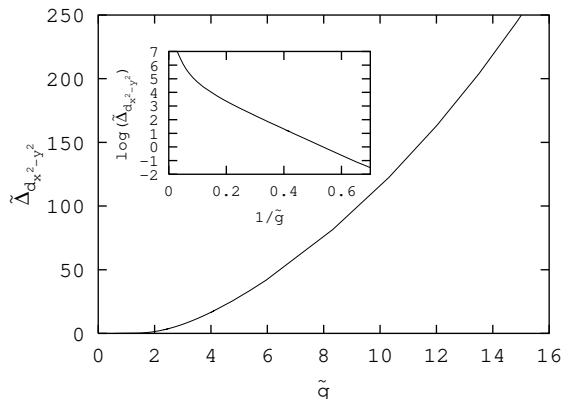}} \par}

\caption{\label{FIG-delta-vs-g-d}Evolution of the amplitude of the gap function
for the \protect\( d_{x^{2}-y^{2}}\protect \) trial ground state,
as a function of \protect\( \tilde{g}\protect \), for fixed \protect\( \tilde{n}=7.5\protect \).}
\end{figure}

\begin{figure}
{\centering \resizebox*{0.9\columnwidth}{!}{\includegraphics{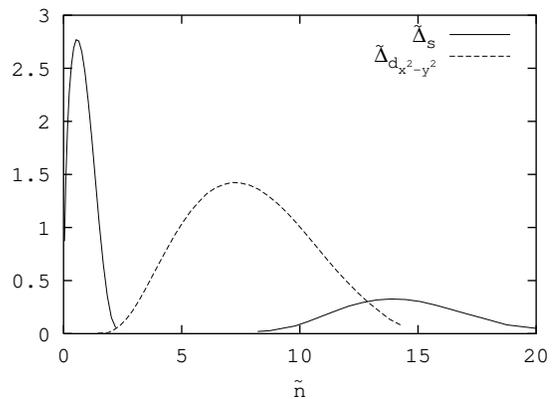}} \par}

\caption{\label{FIG-delta-vs-n}Evolution of the amplitude of the gap function
for the \protect\( s\protect \) and \protect\( d_{x^{2}-y^{2}}\protect \)
trial ground states with increasing value of \protect\( \tilde{n}\protect \),
for fixed \protect\( \tilde{g}=2\protect \). }
\end{figure}

For low values of \( \tilde{g} \), we find that \( \tilde{\mu }\gg \tilde{\Delta }_{s} \)
(\( \tilde{\mu }\gg \tilde{\Delta }_{d_{x^{2}-y^{2}}} \)). This is
the usual weak-coupling condition \cite{Nozieres-SchmittRink-85}
characterising the BCS limit, and consequently the numerical results
display the usual generalised BCS law \cite{Anderson-Brinkman-74}
\( \tilde{\Delta }_{s}\propto \exp \left\{ 1/NK_{0}\right\} \, (\tilde{\Delta }_{d_{x^{2}-y^{2}}}\propto \exp \left\{ 1/NK_{2}\right\} ) \)
where \( N \) is the free-fermion density of states, per spin, per
unit volume. To illustrate this by an example, Fig.~\ref{FIG-delta-vs-g-d}
shows \( \tilde{\Delta }_{d_{x^{2}-y^{2}}} \) vs \( \tilde{g} \)
for constant \( \tilde{n}=7.5 \). In this regime, the non-monotonic
dependence of the effective weak-coupling constant \( K_{l} \) on
the rescaled Fermi vector \( k_{F}r_{0} \) (given in Eq.~(\ref{BCS limit coupling const DSM});
see also Fig.~\ref{FIG-weighting factor}) leads to the similarly
non-monotonic dependence of \( \tilde{\Delta }_{s} \) and \( \tilde{\Delta }_{x^{2}-y^{2}} \)
on \( \tilde{n}=\left( 4/9\pi \right) \left( k_{F}r_{0}\right) ^{3} \)
shown in Fig.~\ref{FIG-delta-vs-n}.%
\footnote{Note the similarity of this behaviour to that of some lattice models
(c.f., for example, the dependence of the critical temperature on
band filling shown in Fig.~18 of Ref.~\onlinecite{Micnas-Ranninger-Robaskiewicz-90}).
}

\begin{figure}
{\centering \resizebox*{0.9\columnwidth}{!}{\includegraphics{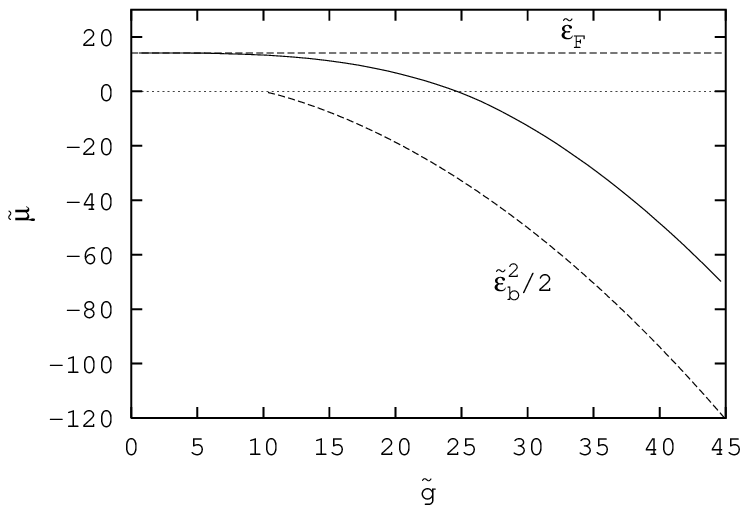}} \par}

\caption{\label{FIG-mu-vs-g}Evolution of the chemical potential for the \protect\( d_{x^{2}-y^{2}}\protect \)
trial ground state, as a function of \protect\( \tilde{g}\protect \),
for fixed \protect\( \tilde{n}=7.5\protect \).}
\end{figure}

On the other hand, for large values of \( \tilde{g} \) we obtain
\( \tilde{\mu }\ll -\tilde{\Delta }_{s} \) (\( \tilde{\mu }\ll -\tilde{\Delta }_{d_{x^{2}-y^{2}}} \)),
which is the opposite strong-coupling condition, corresponding to
the BE limit \cite{Nozieres-SchmittRink-85}. Thus as \( \tilde{g} \)
is increased, while keeping \( \tilde{n} \) constant, \( \tilde{\mu } \)
goes from being approximately independent of \( \tilde{g} \), and
equal to \( \tilde{\varepsilon }_{F} \) (\( \equiv  \) the Fermi
energy \( \varepsilon _{F}\equiv \hbar ^{2}k_{F}^{2}/2m^{*} \) divided
by \( \varepsilon _{0} \)), to having the behaviour\begin{equation}
\label{sc-chem-pot}
\tilde{\mu }\approx \frac{\tilde{\varepsilon }_{b}^{l}}{2}
\end{equation}
where \( \tilde{\varepsilon }_{b}^{l} \) is the binding energy of
the two-body bound state with angular momentum quantum number \( l=0,2 \)
(given analytically in Refs.~\onlinecite{Gottfried-66,Villarroel-98},
for example), divided by \( \varepsilon _{0} \). This evolution of
the chemical potential is represented in Fig.~\ref{FIG-mu-vs-g},
for the \( d_{x^{2}-y^{2}} \) trial ground state (the positive offset
of \( \tilde{\mu } \) above \( \tilde{\varepsilon }_{b}^{l}/2 \)
that can be seen in the graph becomes very small, compared to \( \tilde{\varepsilon }_{b}^{l}/2 \),
only at larger values of \( \tilde{g} \) than those shown; additionally,
it tends to zero as \( \tilde{n}\rightarrow 0 \)).

\begin{figure}
{\centering \resizebox*{0.9\columnwidth}{!}{\includegraphics{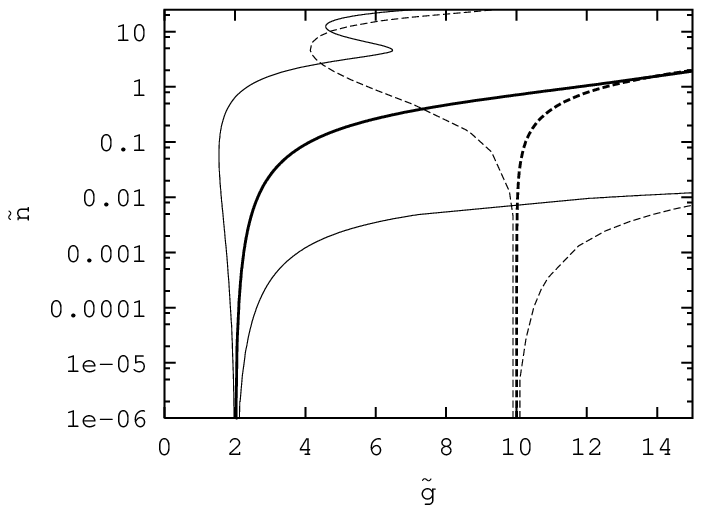}} \par}

\caption{\label{FIG-crossover-chart}{}``Chart'' of the BCS to Bose crossover
for the trial ground state with \protect\( s\protect \) pairing (solid
lines) and the one with \protect\( d_{x^{2}-y^{2}}\protect \) pairing
(dashed lines). The thicker lines are where the chemical potential
goes below the bottom of the band, while the thinner lines give an
indication of the extent of the crossover region (see text).}
\end{figure}

As is well known \cite{Leggett-80,Nozieres-SchmittRink-85} the qualitative
change of the ground state from BCS-like to BE-like behaviour occurs
when the chemical potential goes below the bottom of the band i.e.
\( \tilde{\mu }=0 \). Fig.~\ref{FIG-crossover-chart} shows two
superimposed {}``charts'' of the crossover, for the \( s \) and
\( d_{x^{2}-y^{2}} \) ground states, obtained using this condition.
The charts include also two additional boundaries for each trial ground
state, corresponding to \( \tilde{\mu }=\tilde{\Delta }_{s},\tilde{\Delta }_{d_{x^{2}-y^{2}}} \)
and \( \tilde{\mu }=-\tilde{\Delta }_{s},-\tilde{\Delta }_{d_{x^{2}-y^{2}}} \),
which indicate the extent of the {}``crossover region'' between
the BCS and BE limits. These charts are very similar to the ones presented
in Ref.~\onlinecite{Andrenacci-Perali-Pieri-Strinati-99} for \( s \)-wave
pairing via the the NSR and Gaussian potentials, suggesting that the
density-driven crossover behaviour described in that reference is
generic to continuum models. The main difference that we observe for
\( d_{x^{2}-y^{2}} \)-wave pairing is the enlarged BCS region at
low densities, due to the higher value of the coupling constant required
for a two-body bound state. The oscillations of the boundary between
the BCS and crossover regions at high densities are directly related
to the non-monotonic densitiy-dependence shown in Fig.~\ref{FIG-delta-vs-n}.

\begin{figure}
{\centering \resizebox*{0.9\columnwidth}{!}{\includegraphics{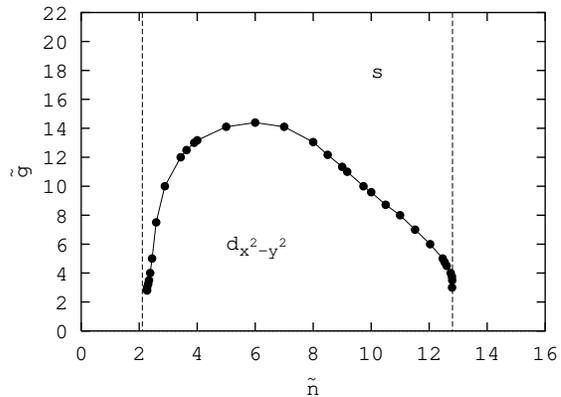}} \par}

\caption{\label{FIG-relative-stability}Phase diagram of the relative stability
of trial ground state with \protect\( s\protect \) and \protect\( d_{x^{2-y^{2}}}\protect \)
pairing symmetry. The dashed lines indicate the position of the phase
boundary in the \protect\( \tilde{g}\rightarrow 0\protect \) limit,
given by Eq.~(\ref{phase transition wc}).}
\end{figure}

Evidently, Fig.~\ref{FIG-delta-vs-n} suggests that at intermediate
densities, at which \( \tilde{\Delta }_{d_{x^{2}-y^{2}}}\gg \tilde{\Delta }_{s} \),
the energy of the trial ground state with \( l=2 \) is lower than
for \( l=0 \). The precise value of the density at which this breaking
of the rotational symmetry takes place, and the higher value at which
the symmetry is restored, are given, in the limit of small \( \tilde{g} \),
by the first two positive solutions of the following equation: \begin{equation}
\label{phase transition wc}
j_{0}\left( \tilde{k}_{F}\right) ^{2}=j_{2}\left( \tilde{k}_{F}\right) ^{2}
\end{equation}
where \( \tilde{k}_{F}\equiv k_{F}r_{0} \). These can be determined
from Fig.~\ref{FIG-weighting factor}. On the other hand, for large
\( \tilde{g} \) the system is always in the BE regime, in which the
energy (\ref{rescaled-internal-energy-equation}) takes the form\begin{equation}
\label{sc-energy}
\tilde{\mathcal{U}}=\frac{1}{2}\tilde{n}\tilde{\varepsilon }_{b}^{l}-\frac{3}{4}\tilde{g}\tilde{n}^{2}
\end{equation}
 Since the Hartree term \( -3/4\tilde{g}\tilde{n}^{2} \) is independent
of \( l \), at first sight this equation suggests that the \( l=0 \)
trial ground state, for which \( \tilde{\varepsilon }_{b}^{l} \)
is lower \cite{Landau-Lifshitz-QM}, must have lower energy, however
note that in general \( \lim \left( \tilde{\mathcal{U}}_{s}-\tilde{\mathcal{U}}_{d_{x^{2}-y^{2}}}\right) \neq \lim \tilde{\mathcal{U}}_{s}-\lim \tilde{\mathcal{U}}_{d_{x^{2}-y^{2}}} \)
(where \( \tilde{\mathcal{U}}_{s} \) and \( \tilde{\mathcal{U}}_{d_{x^{2}-y^{2}}} \)
are the energies of the two trial ground states and the limit refers
to taking \( \tilde{\mu }\ll -\tilde{\Delta }_{s},-\tilde{\Delta }_{d_{x^{2}-y^{2}}} \)
in Eq.~(\ref{rescaled-internal-energy-equation})). In fact there
is an additioanl positive contribution to the energy, similar to the
positive offset of the chemical potential, with respect to \( \varepsilon _{b}^{l}/2 \),
seen in Fig.~\ref{FIG-mu-vs-g}, which does not appear in Eq.~(\ref{sc-energy})
because it varies slowly with \( \tilde{g} \) and therefore becomes
negligible for sufficiently large \( \tilde{g} \) (just like the
offset of \( \tilde{\mu } \)). This repulsion is different for pairs
with different internal structures, and so it is only in the \( \tilde{n}\rightarrow 0 \)
limit which Eq.~(\ref{sc-energy}) allows us to conclude that the
\( l=0 \) state is preferred at high \( \tilde{g} \). At finite
densities, the energies have to be evaluated numerically. Nevertheless
the result, shown in Fig.~\ref{FIG-relative-stability}, confirms
our expectations: between the two densities given by Eq.~(\ref{phase transition wc})
the \( d_{x^{2}-y^{2}} \) trial ground state is more stable, thus
breaking the rotational symmetry of the system, but only for relatively
small values of the coupling constant. As \( \tilde{g} \) is made
larger, the range of densities in which this symmetry is broken becomes
progressively smaller until, above some critical value of \( \tilde{g}\sim 14 \),
the system prefers the \( s \) state at all densities. Thus the region
in parameter space in which the rotational symmetry is broken is relatively
small. In particular, it is confined to the BCS side of the crossover
diagram i.e. \( \tilde{\mu }>0 \) everywhwere inside the \( d_{x^{2}-y^{2}} \)
region.%
\footnote{Note that the fact that the phase transition at these quantum critical
points is of first-order is due to our choice of trial ground states,
Eqs.~(\ref{s-wave-gf},\ref{dxmy2-wave-gf}). Had we allowed for
mixing of the \( s \) and \( d_{x^{2}-y^{2}} \) order parameters,
we would presumably have found two second-order (instead of one first-order)
phase transitions, in analogoy with the similar results in Ref.~\onlinecite{Jon-James-2000}.
}

\section{\label{SEC-Tc}Critical temperature}

Unlike the theory of the ground state that of the equilibrium phase
at finite temperatures does not follow from the usual BCS theory when
the superconducting instability corresponds to the BE condensation
of {}``preformed pairs'' (PP). To describe such situation one must
go beyond the mean-field theory and include fluctuations. This is
most readily done within the framework of a path integral representation
of the partition function \( Z \) \cite{Nagaosa-99,Negele-Orland-88}.
We shall now proceed following this approach and keeping only the
lowest significant corrections to the mean field theory. Namely, we
start with a Grassman path-integral representation of \( Z \) for
the electrons, implement the usual Hubbard-Stratonovich transformation
\cite{Negele-Orland-88,Nagaosa-99,Hubbard-59} to a functional integral
over a complex pairing field \( \Delta  \) and, finally, expand the
effective action for the fluctuations, \( S_{b}\left[ \Delta ^{*},\Delta \right]  \),
about the saddle point of the functional integral above \( T_{c} \)
to quadratic (Gaussian) order \cite{Zwerger-92,Alexandrov-Rubin-93,SaDeMelo-Randeria-Engelbrecht-93,Zwerger-97,Jon-James-2000}.
This is a well-tied approximation for the problem at hand \cite{Randeria-95}
and therefore suitable for studying the effects of pairing fluctuations
on \( T_{c} \) in our particular model. In short, using Eqs.~(\ref{angle decomp of V},\ref{V separable DSM})
to write the Hamiltonian as\begin{eqnarray}
\hat{H}-\mu \hat{N} & = & \sum _{\mathbf{k},\sigma }\varepsilon _{\mathbf{k}}\hat{c}_{\mathbf{k},\sigma }^{+}\hat{c}_{\mathbf{k},\sigma }+\sum _{l,m,\mathbf{q}}\frac{V_{l}}{L^{3}}\hat{b}^{+}_{l,m,\mathbf{q}}\hat{b}_{l,m,\mathbf{q}}\label{our H in k-space separated} 
\end{eqnarray}
 (where \( L^{3} \) is the sample volume and \( V_{l}\equiv \left( -1\right) ^{l+1}g\left( 4\pi r_{0}\right) ^{2} \))
suggests that we introduce bosonic fields \( \Delta _{l,m,\mathbf{q}}\left( \omega _{\nu }\right)  \)
conjugate to the operators\begin{eqnarray}
\hat{b}^{+}_{l,m,\mathbf{q}} & \equiv  & \sum _{\mathbf{k}}j_{l}\left( \left| \mathbf{k}\right| r_{0}\right) Y_{l,m}\left( \hat{\mathbf{k}}\right) \hat{c}^{+}_{\mathbf{q}/2+\mathbf{k}\uparrow }\hat{c}^{+}_{\mathbf{q}/2-\mathbf{k}\downarrow }\label{b+ operator} \\
\hat{b}_{l,m,\mathbf{q}} & \equiv  & \sum _{\mathbf{k}}j_{l}\left( \left| \mathbf{k}\right| r_{0}\right) Y^{*}_{l,m}\left( \hat{\mathbf{k}}\right) \hat{c}_{\mathbf{q}/2-\mathbf{k}\downarrow }\hat{c}_{\mathbf{q}/2+\mathbf{k}\uparrow }\label{b operator} 
\end{eqnarray}
which evidently create and annihilate, respectively, a pair with total
momentum \( \mathbf{q} \) and angular momentum quantum numbers \( l,m \).
As usual, the momentum and frequency dependence of the fields captures
the dynamics of the bosonic degrees of freedom. The additional \( l,m \)-dependence
reflects the fact that our explicit interaction potential can bind
pairs with different internal structures. Obviously the \( \Delta _{l,m} \)
of the previous section correspond to a homogeneous, stationary configuration
of the fields, \( \Delta _{l,m,\mathbf{q}}\left( \omega _{\nu }\right) \equiv \delta _{\mathbf{q},0}\Delta _{l,m} \). 

Proceeding in the usual way \cite{Zwerger-92,Alexandrov-Rubin-93,SaDeMelo-Randeria-Engelbrecht-93,Zwerger-97,Jon-James-2000}
we obtain\begin{eqnarray}
S_{b}\left[ \Delta ^{*},\Delta \right]  & = & \beta \sum _{\mathbf{q}\nu }\sum _{l,l'}\sum _{m,m'}\Delta ^{*}_{l,m,\mathbf{q}}\left( \omega _{\nu }\right) \times \nonumber \\
 &  & \times \Gamma ^{-1}_{l,m,l',m'}\left( \mathbf{q},i\omega _{\nu }\right) \Delta _{l',m',\mathbf{q}}\left( \omega _{\nu }\right) \label{Sb} 
\end{eqnarray}
where the sum on \( l,l' \) extends only over values of the angular
momentum quantum number with the same parity (both even or both odd),
\( \beta \equiv 1/k_{B}T \) is the inverse temperature and the \( \omega _{\nu }\equiv 2\nu \pi /\beta  \)
are bosonic Matsubara frequencies. To further simplify the problem,
and facilitate the discussion of the BE limit, we follow the procedure
employed by Zwerger and coworkers \cite{Zwerger-92,Zwerger-97} to
write a low-frequency, low-momentum expansion of the inverse propagator
for the preformed pairs:\begin{eqnarray}
\Gamma ^{-1}_{l,m,l',m'}\left( \mathbf{q},i\omega _{\nu }\right) /\left[ d_{l}\left( \beta ,\mu \right) d_{l'}\left( \beta ,\mu \right) \right] ^{1/2} & = & \nonumber \\
 &  & \hspace {-6.5cm}\left( -i\omega _{\nu }+\sum _{i=x,y,z}\frac{\hbar ^{2}\mathbf{q}_{i}^{2}}{2m_{l,m,l,m}^{b,ii}\left( \beta ,\mu \right) }-\mu ^{b}_{l}\left( \beta ,\mu \right) \right) \delta _{l,l'}\delta _{m,m'}\nonumber \\
 &  & \hspace {-6.5cm}+\left( \sum _{i,j=x,y,z}\frac{\hbar ^{2}\mathbf{q}_{i}\mathbf{q}_{j}}{2m_{l,m,l',m'}^{b,ij}\left( \beta ,\mu \right) }\right) \left( 1-\delta _{l,l'}\delta _{m,m'}\right) \label{Gamma^{-}1 bosons} 
\end{eqnarray}
Thus after appropriate rescaling of the bosonic fields the known functions
\( m_{l,m,l',m'}^{b,ij}\left( \beta ,\mu \right)  \) and \( \mu ^{b}_{l}\left( \beta ,\mu \right)  \)
play the r\^ole of effective boson masses and chemical potentials,
as in Refs.~\onlinecite{Nozieres-SchmittRink-85,Zwerger-92,SaDeMelo-Randeria-Engelbrecht-93,Zwerger-97},
for example. These two functions, and the rescaling factor \( d_{l}\left( \beta ,\mu \right)  \),
are given in the appendix. 

Note the different chemical potentials for bosons with different values
of the angular momentum. Moreover, the anisotropic dispersion relation
given by \( m_{l,m,l',m'}^{b,ij}\left( \beta ,\mu \right)  \) can
describe not only the {}``rigid'' propagation of a boson without
changing its internal state, but also changes in its internal angular
momentum through the off-diagonal terms, with \( l,m\neq l',m' \).
However in the BE limit, which as usual corresponds to \( \mu \beta \rightarrow -\infty  \),
we have \( m_{l,m,l',m'}^{b,ij}\left( \beta ,\mu \right) \rightarrow \infty  \)
for \( l,m\neq l',m' \), and therefore in what follows we shall ignore
these off-diagonal terms (for \( l,m=l',m' \), on the other hand,
we recover the expected behaviour \cite{Nozieres-SchmittRink-85,Zwerger-92,SaDeMelo-Randeria-Engelbrecht-93,Zwerger-97}:
\( m_{l,m}^{b,i}\left( \beta ,\mu \right) \equiv m_{l,m,l,m}^{b,ii}\left( \beta ,\mu \right) \rightarrow 2m^{*} \)). 

As usual, \( T_{c} \) is determined by the BE condensation condition
\( \mu _{l}^{b}\left( \beta ,\mu \right) =0 \). This gives a different
critical temperature for each value of \( l \). On the other hand
\( T_{c} \) is degenerate in \( m=-l\ldots l \), as in BCS theory
\cite{Balian-64}. Since the present method can only describe an instability
of the normal state, our philosophy will be to compute the \( l=0 \)
and \( l=2 \) critical temperatures and then take the highest of
the two as the critical temperature of the system. Moreover we will
assume that, near the critical temperature, only fluctuations with
the appropriate value of \( l \) have to be taken into account. This
is only adequate if the \( l=0 \) and \( l=2 \) critical temperatures
differ considerably, which as we shall see is the case in the BE limit.
Of course in the opposite, BCS limit the fluctuations can be neglected
completely.

Under the above assumptions the {}``\( T_{c} \)'' and {}``density''
equations take the form \begin{equation}
\label{rescaled-tc-eq}
\frac{1}{\tilde{g}}=\left( -1\right) ^{l}\frac{2}{\pi }\int _{0}^{\infty }d\left| \tilde{\mathbf{k}}\right| \left| \tilde{\mathbf{k}}\right| ^{2}j_{l}\left( \left| \tilde{\mathbf{k}}\right| \right) ^{2}\frac{1-2f\left( \tilde{\beta }\tilde{\varepsilon }_{\tilde{\mathbf{k}}}\right) }{2\tilde{\varepsilon }_{\tilde{\mathbf{k}}}}
\end{equation}
\begin{equation}
\label{density-eq-rescaled}
\tilde{n}=\tilde{n}_{0}+\sum _{m}\delta \tilde{n}_{l,m}
\end{equation}
 where \( \tilde{n}_{0}=\frac{4}{3\pi }\int _{0}^{\infty }d\left| \tilde{\mathbf{k}}\right| \left| \tilde{\mathbf{k}}\right| ^{2}f\left( \tilde{\beta }\tilde{\varepsilon }_{\tilde{\mathbf{k}}}\right)  \)
is the density of fermions that are unpaired above \( T_{c} \) and
the additional contribution coming from Gaussian fluctuations is made
up of terms of the form \begin{equation}
\label{deltanlm-rescaled}
\delta \tilde{n}_{l,m}=\frac{4}{3\pi }\int _{0}^{\infty }d\left| \tilde{\mathbf{q}}\right| \left| \tilde{\mathbf{q}}\right| ^{2}g\left( \tilde{\beta }\frac{\left| \tilde{\mathbf{q}}\right| ^{2}}{2}\right) \tilde{w}\left( \left| \tilde{\mathbf{q}}\right| \right) 
\end{equation}
which correspond to fermions bound in PP with angular momentum given
by \( l,m \). The notations \( f\left( x\right)  \) and \( g\left( x\right)  \)
have been used for the Fermi and Bose distributions functions, respectively.
The {}``weight'' \( \tilde{w}\left( \left| \tilde{\mathbf{q}}\right| \right)  \)
is given by \begin{eqnarray}
\tilde{w}\left( \left| \tilde{\mathbf{q}}\right| \right)  & \equiv  & \left( \prod _{i=1}^{3}\tilde{m}_{l,m}^{b,i}\right) ^{1/2}\times \nonumber \\
 &  & \hspace {-2cm}\times \left[ 1+\frac{1}{\tilde{d}_{l}}\left( \frac{\tilde{d}_{l}'}{\tilde{d}_{l}}\tilde{\lambda }_{l}-\tilde{\lambda }_{l}'\right) \frac{1}{3}\left( \sum _{i}\kappa ^{i}_{l,m}\tilde{m}_{l,m}^{b,i}\right) \frac{\left| \tilde{\mathbf{q}}\right| ^{2}}{2}\right] \label{rescaled-w} 
\end{eqnarray}
where the dimensionless function \( \tilde{\lambda }_{l}\left( \tilde{\beta },\tilde{\mu }\right)  \)
and the factor \( \kappa _{l,m}^{i} \) are defined in the appendix
and each {}``primed'' represents differentiation with respect to
\( \tilde{\mu } \). This weight becomes unity in the BE limit \( \mu \beta \rightarrow -\infty  \),
corresponding to bosons of mass \( 2m^{*} \) each.

\begin{figure}
{\centering \resizebox*{0.9\columnwidth}{!}{\includegraphics{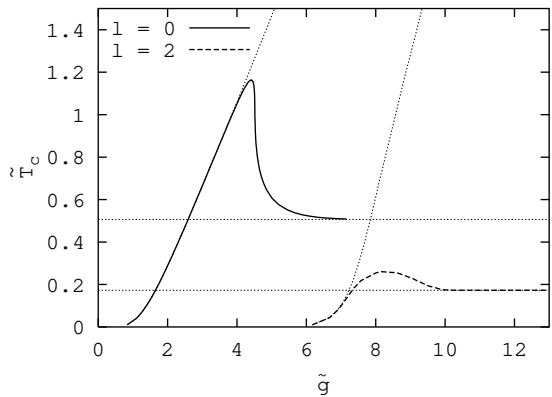}} \par}

\caption{\label{FIG-tcv_vs_g}The critical temperature for an instability
to a superconducting state with \protect\( l=0\protect \) (solid
line) and \protect\( l=2\protect \) (dashed line), as a function
of \protect\( \tilde{g}\protect \), for fixed \protect\( \tilde{n}=0.5\protect \).
The increasing dotted lines are obtained by neglecting the contribution
of Gaussian fluctuations to the total density, i.e. the second term
on the right-hand side of Eq.~(\ref{density-eq-rescaled}), while
the constant dotted lines are the BE condensation temperature given
in Eq.~(\ref{TcBE}).}
\end{figure}

Numerical solution of the self-consistency equations (\ref{rescaled-tc-eq},\ref{density-eq-rescaled})
for \( \tilde{\beta } \) and \( \tilde{\mu } \), at the relatively
low value of the density \( \tilde{n}=0.5 \), shows the expected
\cite{Nozieres-SchmittRink-85,Zwerger-92,SaDeMelo-Randeria-Engelbrecht-93,Zwerger-97}
smooth evolution between the BCS and BE limits, analogous to the one
seen in the ground state: see Fig.~\ref{FIG-tcv_vs_g}. In particular,
we find that the critical temperature for angular momentum quantum
number \( l \) is \begin{equation}
\label{TcMF}
\tilde{T}_{c}^{l}\propto \exp \left\{ 1/NK_{l}\right\} 
\end{equation}
 for small \( \tilde{g} \) but saturates to a constant value given
by \begin{equation}
\label{TcBE}
k_{B}T_{c}\approx 3.315\frac{\hbar ^{2}}{2m^{*}}\left[ \frac{n}{2\left( 2l+1\right) }\right] ^{2/3}
\end{equation}
in the large-\( \tilde{g} \) limit. \foreignlanguage{british}{}This
asymptotic behaviour follows quite generally from the self-consistency
Eqs.~(\ref{rescaled-tc-eq},\ref{density-eq-rescaled}).

Notably, Eq.~(\ref{TcBE}) differs from the standard result (\ref{TcBE Ogg})
in the presence of the degeneracy factor \( 1/\left( 2l+1\right)  \)
multiplying the density of bosons \( n/2 \). For an instability to
an \( s \)-wave superconducting state, with \( l=0 \), Eq.~(\ref{TcBE})
reduces to Eq.~(\ref{TcBE Ogg}) and thus our result for the DSM
coincides with those obtained earlier for models featuring the NSR
\cite{Nozieres-SchmittRink-85} and contact \cite{Zwerger-92,SaDeMelo-Randeria-Engelbrecht-93,Haussmann-93}
potentials. On the other hand, the degeneracy of the \( l=2 \) bound
state means that, at \( T_{c} \), five Bose gases condense simultaneously,
but independently, leading to a much lower critical temperature. This
is in contrast with the result for the anisotropic potential of Ref.~\onlinecite{Zwerger-97}.
On the basis of this we conclude that the \( l=0 \) state always
has higher critical temperature in the BE limit.

\begin{figure}
{\centering \resizebox*{0.9\columnwidth}{!}{\includegraphics{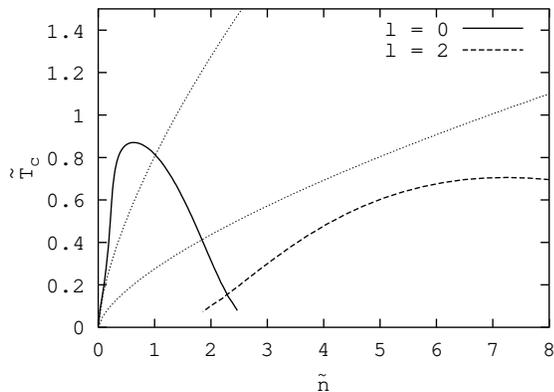}} \par}

\caption{\label{FIG-tc_vs_n}The critical temperature for an instability to
a superconducting state with \protect\( l=0\protect \) (solid line)
and \protect\( l=2\protect \) (dashed line), as a function of \protect\( \tilde{n}\protect \),
for fixed \protect\( \tilde{g}=3.5\protect \). The dotted lines represent
the BE condensation temperatures given by Eq.~(\ref{TcBE}).}
\end{figure}

On the other hand, for small values of \( \tilde{g} \) (\( \sim 3.5 \))
we find a non-monotonic density-dependence of the \( s \)- and \( d \)-wave
critical temperatures similar to the one that we described for the
amplitude of the gap function in the respective trial ground states:
see Fig.~\ref{FIG-tc_vs_n}. In particular, note that there is an
intermediate range of densities for which the \( d \)-wave critical
temperature is the highest.

\begin{figure}
{\centering \resizebox*{0.9\columnwidth}{!}{\includegraphics{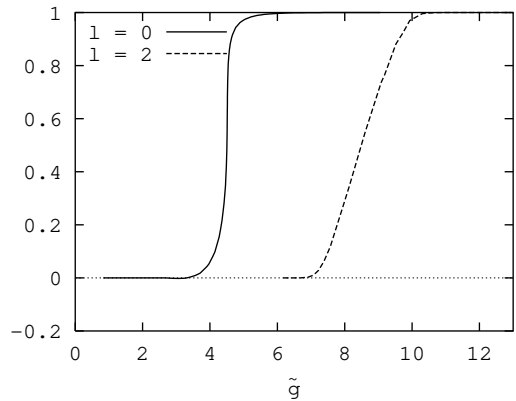}} \par}

\caption{\label{FIG-dnc_both_0p5}The fraction of fermions that are bound
into PP just above \protect\( T_{c}\protect \) for an instability
to a superconducting state with \protect\( l=0\protect \) (solid
line) and \protect\( l=2\protect \) (dashed line), as a function
of \protect\( \tilde{g}\protect \), at the density of Fig.~\ref{FIG-tcv_vs_g}.}
\end{figure}

\begin{figure}
{\centering \resizebox*{0.9\columnwidth}{!}{\includegraphics{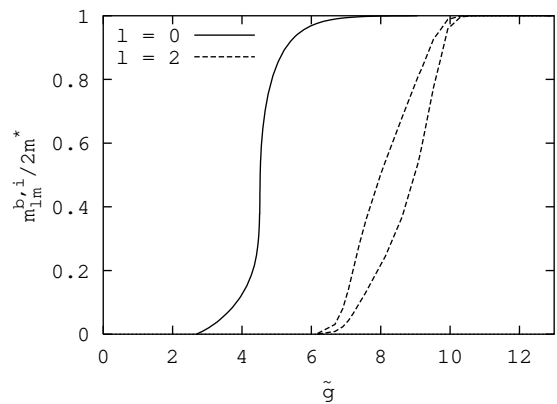}} \par}

\caption{\label{FIG-mbc_both_0p5}The effective mass of the PP existing just
above \protect\( T_{c}\protect \) for an instability to a superconducting
state with \protect\( l=0\protect \) (solid line) and \protect\( l=2\protect \)
(dashed line), as a function of \protect\( \tilde{g}\protect \),
for the density of Fig.~\ref{FIG-tcv_vs_g}. For the case of an instability
to a superconducting state with \protect\( l=2\protect \), only the
heaviest and the lightest of the masses \protect\( m_{l,m}^{b,i}\protect \)
(corresponding to \protect\( i=z\protect \) and \protect\( \left| m\right| =2\protect \)
and \protect\( 0\protect \), respectively) have been plotted.}
\end{figure}

As expected \cite{Nozieres-SchmittRink-85,Zwerger-92,Zwerger-97},
the evolution from the BCS to the Bose limits is also evidenced in
the fraction of fermions that are bound in PP just above \( T_{c} \),
, \( \delta n/n \) (with \( \delta \tilde{n}\equiv \sum _{m}\delta \tilde{n}_{l,m} \)),
and in the effective mass of such PP, \( m_{l,m}^{b,i} \) (see Figs.~\ref{FIG-dnc_both_0p5}
and \ref{FIG-mbc_both_0p5}, respectively): both are negligible for
small \( \tilde{g} \), while in the large-\( \tilde{g} \) limit
we have \( \delta n\approx n \) and \( m_{l,m}^{b,i}\approx 2m^{*} \).

\begin{figure}
{\centering \resizebox*{0.9\columnwidth}{!}{\includegraphics{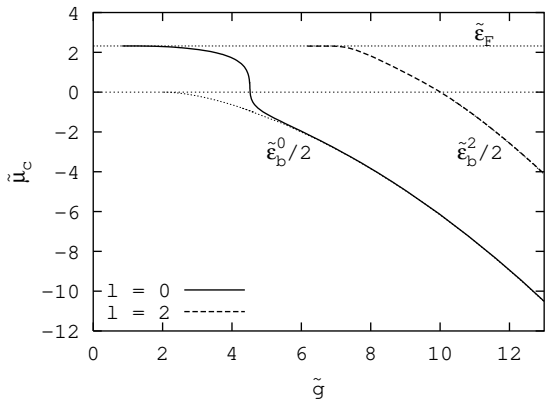}} \par}

\caption{\label{FIG-mc_both}The critical value of the chemical potential
for an instability to a superconducting state with \protect\( l=0\protect \)
(\protect\( s\protect \)-wave, solid line) and \protect\( l=2\protect \)
(\protect\( d\protect \)-wave, dashed line), as a function of \protect\( \tilde{g}\protect \),
for fixed \protect\( \tilde{n}=0.5\protect \). The dotted lines indicate
the Fermi energy \protect\( \tilde{\varepsilon }_{F}\protect \) and
the two-body binding energies per particle \protect\( \tilde{\varepsilon }_{b}^{l}/2\protect \).}
\end{figure}

Like any theory based on a Gaussian expansion, the present one displays
a non-monotonic behaviour of the critical temperature as a function
of \( \tilde{g} \) in the intermediate regime (see Fig.~\ref{FIG-tcv_vs_g}).
Such enhancement \cite{Nozieres-SchmittRink-85,Zwerger-92,SaDeMelo-Randeria-Engelbrecht-93}
is not present \cite{Haussmann-94} in the self-consistent theory
\cite{Haussmann-93} due to Haussmann, suggesting that it is an artifact.
It can be understood in terms of Eq.~(\ref{TcBE}) and Fig.~\ref{FIG-mbc_both_0p5}
as a result of the PP getting lighter (\( m_{l,m}^{b,i}\alt 2m^{*} \))
as the value of \( \tilde{g} \) is reduced. In the self-consistent
theory, at least for a model based on the contact potential, repulsive
interactions between the PP overcompensate for this, leading to a
monotonic dependence of \( T_{c} \) on the coupling constant \cite{Haussmann-94}
(for a more advanced treatment of these repulsive forces see Ref.~\onlinecite{Pieri-Strinati-00};
alternative methods to describe phase fluctuations and strong pairing
correlations in superconductors are described in Refs.~\onlinecite{Gyorffy-Staunton-Stocks-91}
and \onlinecite{Chen-Kosztin-Levin-00}). Such interactions are completely
neglected in the present treatment as is evidenced for example in
Fig.~\ref{FIG-mc_both} which shows the evolution of the chemical
potential, lacking a positive offset like the one we found in the
ground state (compare the \( l=2 \) curve of Fig.~\ref{FIG-mc_both}
to Fig.~\ref{FIG-mu-vs-g}). Moreover, the description of the Gaussian
flctuations afforded by Eq.~(\ref{Gamma^{-}1 bosons}) turns out
to be valid only for densities below the first maximum of \( \tilde{T}_{c} \),
as a function of \( \tilde{n} \). At the maximum, the mass of the
PP (given in Eq.~(\ref{boson-mass}), below) becomes negative, thus
making \( \delta \tilde{n} \) diverge. The present treatment is therefore
only valid at small values of the coupling constant, for which the
fluctuations can be neglected (as in Fig.~\ref{FIG-tc_vs_n}: the
only part of the plot that shows a significant contribution from fluctuations
is at densities well below the first maximum of \( \tilde{T}_{c} \))
or at low densities, which are below the first maximum for all sizeable
values of \( \tilde{g} \) (as in Fig.~\ref{FIG-tcv_vs_g}).

\section{Conclusions}

We have studied exotic pairing in the context of a simple model featuring
fermions in a continuum with an explicit, non-retarded, central interaction
potential \( V\left( r\right)  \): the Delta-Shell Model (DSM). Its
novel feature is that the interaction is attractive only at some finite
distance \( r_{0} \). Because of this it provides, to our knowledge,
the first explicit example of BCS pairing with angular momentum quantum
number \( l>0 \) \textit{via} a spherically symmetric (central) attraction.
By using a standard functional integral approach, we have studied
this breaking of the rotational symmetry in relation to the BCS to
Bose crossover. 

By considering two trial ground states, with \( s \) and \( d_{x^{2}-y^{2}} \)
symmetries, we have found that \emph{the ground state with broken
rotational symmetry is separated from the BE regime by a quantum phase
transition}, in which the symmetry of the superconducting order parameter
is increased. This is due to the higher \emph{energy} of two-body
bound states with \( l>0 \), and so it can be extrapolated to any
central interaction potential. More generally, for other models (such
as those in which the single-particle dispersion relation and the
interaction potential are anisotropic), our analysis suggests that
a two-body ground state with \( d_{x^{2}-y^{2}} \) symmetry is required
in order for pairing to take that form in the BE regime. Such scenario
is realised, for example, in a lattice model with nearest neighbour
(n.n.) attraction and large next nearest neighbour (n.n.n.) hopping
\cite{Bak-Micnas-00}. 

Similarly, the critical temperature for superconductivity with angular
momentum quantum number \( l=0 \) is found to be higher than for
\( l=2 \) in the BE limit (of strong coupling \emph{and} low densities).
However, interestingly, this is due not to the higher energy, but
to the related higher \emph{degeneracy} of the two-body bound state
with \( l=2 \). Thus together these two observations place severe
constraints on any interaction potential \( V\left( \mathbf{r}\right)  \)
leading to pairing with \( l>0 \) in the BE limit. 

In our model, the rotational symmetry-breaking is a direct consequence
of a non-monotonic dependence of the superconducting properties on
the fermion density which is present only in the BCS regime. Such
rise-and-falls can be understood in terms of the oscillatory form
of the {}``gap function'' in \( \mathbf{k} \)-space, whose frequency
is \( \sim r_{0}^{-1} \), and presumably they are generic to interaction
potentials that are attractive predominantly at a finite distance.
In Refs.~\onlinecite{Quintanilla-Gyorffy-00,Quintanilla-Gyorffy-02}
the possible implications of our model to cuprate superconductors,
on the basis of the similar behaviour observed in the doping-dependence
of the supercondcuting gap and the critical temperature, were discussed.
A similar rise-and-fall has been known for some time in nuclei (see
Ref.~\onlinecite{Sarriguren-MoyaDeGuerra-Lombardo-Schuck-Schulze-01},
for example). Of particular interest, in connection with recent theoretical
speculations on superfluidity in magnetically trapped gases of fermionic
atoms \cite{Bruun-Castin-Dum-Burnett-99,Combescot-01,Holland-Kokkelmans-Chiofalo-Walser-01,Ohashi-Griffin-02}
(for an informal review and further references see Ref.~\onlinecite{Jin-02}),
is the possibility that the present mechanism would lead to exotic
pairing for sufficiently high densities in these systems. Interestingly,
because the change from \( s \)- to \( d \)-wave pairing is a quantum
phase transition, it can take place at arbitrarily low temperatures.
On the other hand, the density would have to be raised until \( r_{s}\sim r_{0} \)
(where, in order to achieve a phenomenological description of the
inter-atomic potential, \( r_{0} \) may be taken to be roughly the
size of a diatomic molecule). 

In relation to possible future work, we end by noting that we have
described the ground state of the DSM in terms of a homogeneous saddle
point, and we have only taken into account pairing fluctuations around
that saddle point. In principle, by performing a more general Hubbard-Stratonovich
transformation, including additional fields associated with the density
(not just the pairing amplitude), one could study the effect of density
fluctuations as well as the possibility of phase separation through
a first-order gas-liquid phase transition: physically, one expects
that the attraction at a \emph{finite} distance could favour, in addition
to the pairing with \( l>0 \) which we have considered here, the
formation of clusters of more than two particles (as in lattice models
with nearest-neighbour attraction \cite{Lin-Hirsch-86}, and unlike
those with on-site attraction \cite{Nozieres-SchmittRink-85}; for
a discussion of the similar phenomenon of {}``quartetting'' see
Ref.~\onlinecite{Ropke-Schnell-Schuck-Nozieres-98}; see also footnote
19 of Ref.~\onlinecite{Pieri-Strinati-00}). Evidently, this would
be very interesting in the light of recent discussions of inhomogeneity
in cuprate superconductors \cite{Burgy-Mayr-MartinMayor-Moreo-Dagotto-01}.

\begin{acknowledgments}
We thank K.~Capelle for detailed comments on an earlier manuscript
and A.S.~Alexandrov, N.~Andrenacci, N.W. Ashcroft, J.M.F.~Gunn,
P.~Howell and H.M.~Kleinert for useful discussions. JQ acknowledges
financial support from the TMR programme (EU; contract No.~ERBFMBICT983194),
FAPESP (Brazil; process No.~01/1046 1-8) and CNPq (Brazil).
\end{acknowledgments}
\appendix

\section{Expansion of the inverse bosonic propagator}

\begin{figure}
{\centering \resizebox*{0.9\columnwidth}{!}{\includegraphics{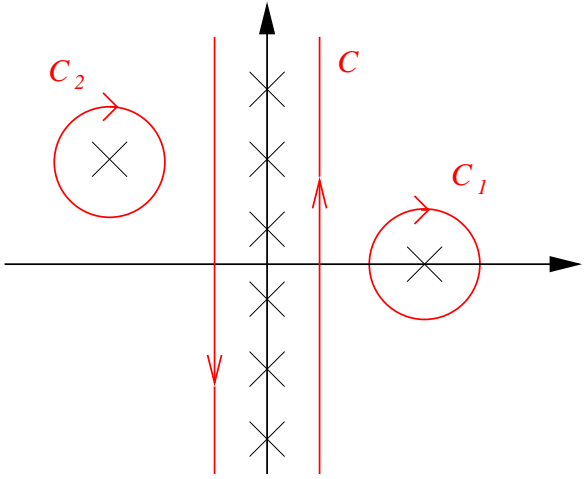}} \par}

\caption{\label{FIG-poles}The contours used to perform the summation over
the fermionic Matsubara frequencies \protect\( \omega _{n}\protect \)
in Eq. (\ref{intermediate-eq-in-the-appendix}).}
\end{figure}

The inverse bosonic propagator in Eq.~(\ref{Sb}) is given by \begin{eqnarray}
\Gamma ^{-1}_{l,m,l',m'}\left( \mathbf{q},i\omega _{\nu }\right)  & = & \nonumber \\
 &  & \hspace {-3cm}\frac{L^{3}}{g}\delta _{l,l'}\delta _{m,m'}-\frac{1}{\beta }\sum _{n}\sum _{\mathbf{k}}i^{l'-l}\Lambda _{l,m,l',m'}\left( \mathbf{k}\right) \times \nonumber \\
 &  & \hspace {-3cm}\times G_{0}\left( \frac{\mathbf{q}}{2}+\mathbf{k},i\omega _{n}\right) G_{0}\left( \frac{\mathbf{q}}{2}-\mathbf{k},i\omega _{\nu }-i\omega _{n}\right) \label{intermediate-eq-in-the-appendix} 
\end{eqnarray}
 where \( G_{0}\left( \mathbf{k},i\omega _{n}\right) \equiv \left( i\omega _{n}-\varepsilon _{\mathbf{k}}\right) ^{-1} \)
(with \( \varepsilon _{\mathbf{k}}\equiv \varepsilon _{0}\tilde{\varepsilon }_{\tilde{\mathbf{k}}} \))
is the free fermion Green's function and the \( \omega _{n}\equiv \left( 2n+1\right) \pi /\beta  \)
are fermionic Matsubara frequencies. The derivation, starting from
Eq.~(\ref{our H in k-space separated}), is entirely analogous to
that of the similar expression in Ref.~\onlinecite{Zwerger-92}, for
example. In Ref.~\onlinecite{SaDeMelo-Randeria-Engelbrecht-93}, the
full frequency dependence of \( \Gamma ^{-1}\left( \mathbf{q},i\omega _{\nu }\right)  \)
was taken into account to obtain the critical temperature of a model
featuring the contact potential. The procedure that we follow here
\cite{Zwerger-92,Zwerger-97} yields the same results in the BCS and
BE limits and a much simpler numerical problem in the crossover regime
(where any theory based on a Gaussian expansion must be regarded as
an interpolation scheme anyway). First we analytically continue the
second Green's function on the right-hand side of Eq.~(\ref{intermediate-eq-in-the-appendix})
with respect to the bosonic Matsubara frequency, \( G_{0}\left( \frac{\mathbf{q}}{2}-\mathbf{k},i\omega _{\nu }-i\omega _{n}\right) \rightarrow G_{0}\left( \frac{\mathbf{q}}{2}-\mathbf{k},w-i\omega _{n}\right) , \)
and then we perform the summation over \( n \). Using the contour
\( C \) in Fig.~\ref{FIG-poles} (which we deform into \( C_{1} \)
and \( C_{2} \)) we obtain\begin{eqnarray}
\Gamma ^{-1}_{l,m,l',m'}\left( \mathbf{q},w\right)  & = & \nonumber \\
 &  & \hspace {-2cm}\frac{L^{3}}{g}\delta _{l,l'}\delta _{m,m'}-\sum _{\mathbf{k}}i^{l'-l}\Lambda _{l,m,l',m'}\left( \mathbf{k}\right) \times \nonumber \\
 &  & \hspace {-2cm}\times \frac{1-f\left( \beta \varepsilon _{\mathbf{q}/2+\mathbf{k}}\right) -f\left( \beta \left[ \varepsilon _{\mathbf{q}/2-\mathbf{k}}-w\right] \right) }{\varepsilon _{\mathbf{q}/2+\mathbf{k}}+\varepsilon _{\mathbf{q}/2-\mathbf{k}}-w}\label{continued inv Gamma} 
\end{eqnarray}
It is now easy to write a low-frequency, low-momentum expansion of
the form \begin{widetext}\begin{equation}
\label{inv Gamma expansion}
\Gamma ^{-1}_{l,m,l',m'}\left( \mathbf{q},i\omega _{\nu }\right) \approx a_{l,m,l',m'}\left( \beta ,\mu \right) -id_{l,m,l',m'}\left( \beta ,\mu \right) \omega _{\nu }+\sum _{i,j=x,y,z}\frac{\hbar ^{2}}{2m^{*}}c^{i,j}_{l,m,l',m'}\left( \beta ,\mu \right) \mathbf{q}_{i}\mathbf{q}_{j}
\end{equation}
 \end{widetext} by simply differentiating with respect to \( \mathbf{q} \)
and \( w \).%
\footnote{Note that Eq.~(\ref{continued inv Gamma}) does not coincide with
the similar expression in Ref.~\onlinecite{SaDeMelo-Randeria-Engelbrecht-93},
obtained by analytical continuation of the inverse bosonic propagator:
\( \Gamma ^{-1}_{l,m,l',m'}\left( \mathbf{q},i\omega _{\nu }\right) \rightarrow \Gamma ^{-1}_{l,m,l',m'}\left( \mathbf{q},w\right)  \).
Apart from the model-specific features, which are our main concern
here, it differs also in the presence of the continuous variable \( w \)
in the argument of one of the Fermi distribution functions. At the
bosonic Matsubara frequencies \( w=i\omega _{\nu } \), we can write
\( f\left( \beta \left[ \varepsilon _{\mathbf{q}/2-\mathbf{k}}-i\omega _{\nu }\right] \right) =f\left( \beta \varepsilon _{\mathbf{q}/2-\mathbf{k}}\right)  \)
and therefore both expressions are identical when the full frequency-dependence
of \( \Gamma ^{-1}_{l,m,l',m'}\left( \mathbf{q},i\omega _{\nu }\right)  \)
is taken into account. However the expression in Ref.~\onlinecite{SaDeMelo-Randeria-Engelbrecht-93}
does not admit a small-frequency expansion of the form (\ref{inv Gamma expansion})
because as is well known \cite{Nozieres-SchmittRink-85,Randeria-95}
it has a branch cut along the real axis that crosses the imaginary
axis whenever \( \mu >0 \). (On the other hand such expression is
the correct starting point for the derivation of a time-dependent
Ginzburg-Landau theory \cite{SaDeMelo-Randeria-Engelbrecht-93}, which
obviously involves an expansion along the \emph{real} axis.) 
} 
We find that the coefficients \( a_{l,m,l',m'} \) and \( d_{l,m,l',m'} \)
are diagonal in \( l,m \), and degenerate in \( m \): \( a_{l,m,l',m'}\left( \beta ,\mu \right) =a_{l}\left( \beta ,\mu \right) \delta _{l,l'}\delta _{m,m'} \)
and \( d_{l,m,l',m'}\left( \beta ,\mu \right) =d_{l}\left( \beta ,\mu \right) \delta _{l,l'}\delta _{m,m'} \).
The dimensionless functions \( \tilde{a}_{l}\left( \tilde{\beta },\tilde{\mu }\right) \equiv r_{0}\varepsilon _{0}L^{-3}a_{l}\left( \beta ,\mu \right)  \)
and \( \tilde{d}_{l}\left( \tilde{\beta },\tilde{\mu }\right) \equiv r_{0}\varepsilon _{0}^{2}L^{-3}d_{l}\left( \beta ,\mu \right)  \)
are given by\begin{eqnarray}
\tilde{a}_{l}\left( \tilde{\beta },\tilde{\mu }\right)  & = & \frac{1}{\tilde{g}}-\left( -1\right) ^{l}\frac{2}{\pi }\int _{0}^{\infty }d\left| \tilde{\mathbf{k}}\right| \left| \tilde{\mathbf{k}}\right| ^{2}\times \label{rescaled-a} \\
 &  & \times j_{l}\left( \left| \tilde{\mathbf{k}}\right| \right) ^{2}\frac{1-2f\left( \tilde{\beta }\tilde{\varepsilon }_{\tilde{\mathbf{k}}}\right) }{2\tilde{\varepsilon }_{\tilde{\mathbf{k}}}}\nonumber \\
\tilde{d}_{l}\left( \tilde{\beta },\tilde{\mu }\right)  & = & \left( -1\right) ^{l}\frac{2}{\pi }\int _{0}^{\infty }d\left| \tilde{\mathbf{k}}\right| \left| \tilde{\mathbf{k}}\right| ^{2}j_{l}\left( \left| \tilde{\mathbf{k}}\right| \right) ^{2}\times \label{rescaled-d} \\
 &  & \times \left\{ \frac{1-2f\left( \tilde{\beta }\tilde{\varepsilon }_{\tilde{\mathbf{k}}}\right) }{4\tilde{\varepsilon }_{\tilde{\mathbf{k}}}^{2}}+\frac{\tilde{\beta }f'\left( \tilde{\beta }\tilde{\varepsilon }_{\tilde{\mathbf{k}}}\right) }{2\tilde{\varepsilon }_{\tilde{\mathbf{k}}}}\right\} \nonumber 
\end{eqnarray}
where we have used the notation \( f'\left( x\right) \equiv df\left( x\right) /dx \)
and the dimensionless inverse temperature is \( \tilde{\beta }\equiv \varepsilon _{0}\beta  \).
The coefficient in \( \mathbf{q}^{2} \) has the form \begin{eqnarray}
c_{l,m,l',m'}^{i,j}\left( \beta ,\mu \right)  & = & \frac{1}{2}d_{l}\left( \beta ,\mu \right) \delta _{l,l'}\delta _{m,m'}\delta _{i,j}\nonumber \\
 &  & +\kappa _{l,m,l',m'}^{i,j}\lambda _{l,l'}\left( \beta ,\mu \right) \label{c-coefficient} 
\end{eqnarray}
 where \( \kappa _{l,m,l',m'}^{i,j}\equiv \int _{\Sigma \left( 1\right) }d^{2}\mathbf{k}Y^{*}_{l,m}\left( \hat{\mathbf{k}}\right) \hat{\mathbf{k}}_{i}\hat{\mathbf{k}}_{j}Y_{l',m'}\left( \hat{\mathbf{k}}\right)  \),
which can be evaluated easily by writing it in terms of the Gaunt
coefficients of relativistic quantum mechanics (see Ref.~\onlinecite{Strange-98},
for example; note in particular that \( \kappa _{l,m,l,m}^{i,j}=0 \)
for all \( i\neq j \)) and, finally, \( \tilde{\lambda }_{l,l'}\left( \tilde{\beta },\tilde{\mu }\right) \equiv r_{0}\varepsilon _{0}\lambda _{l,l'}\left( \beta ,\mu \right)  \)
is given by \begin{eqnarray}
\tilde{\lambda }_{l,l'}\left( \tilde{\beta },\tilde{\mu }\right)  & = & i^{l'-l}\frac{2}{\pi }\int _{0}^{\infty }d\left| \tilde{\mathbf{k}}\right| \left| \tilde{\mathbf{k}}\right| ^{2}j_{l}\left( \left| \tilde{\mathbf{k}}\right| \right) \times \nonumber \\
 &  & \times j_{l'}\left( \left| \tilde{\mathbf{k}}\right| \right) \left| \tilde{\mathbf{k}}\right| ^{2}\frac{\tilde{\beta }^{2}f''\left( \tilde{\beta }\tilde{\varepsilon }_{\tilde{\mathbf{k}}}\right) }{2\tilde{\varepsilon }_{\tilde{\mathbf{k}}}}\label{rescaled-lambda} 
\end{eqnarray}

The integrands of Eqs.~(\ref{rescaled-a}), (\ref{rescaled-d}) and
(\ref{rescaled-lambda}) have no poles on the domain of integration,
and therefore are straight-forward to evaluate numerically. From comparison
of (\ref{inv Gamma expansion}) to (\ref{Gamma^{-}1 bosons}) it is
evident that the masses and chemical potentials of the PP are given
by\begin{eqnarray}
\tilde{\mu }^{b}_{l}\left( \tilde{\beta },\tilde{\mu }\right)  & = & -\tilde{a}_{l}\left( \tilde{\beta },\tilde{\mu }\right) /\tilde{d}_{l}\left( \tilde{\beta },\tilde{\mu }\right) \\
\tilde{m}_{l,m,l',m'}^{b,i,j}\left( \tilde{\beta },\tilde{\mu }\right)  & = & \left[ \delta _{l,l'}\delta _{m,m'}+\frac{2\kappa _{l,m,l',m'}^{i,j}\tilde{\lambda }_{l,l'}}{\left( \tilde{d}_{l}\tilde{d}_{l'}\right) ^{1/2}}\right] ^{-1}\label{boson-mass} 
\end{eqnarray}
where we have introduced the definitions \( \tilde{\mu }^{b}_{l}\left( \tilde{\beta },\tilde{\mu }\right) \equiv \varepsilon _{0}^{-1}\mu ^{b}_{l}\left( \beta ,\mu \right)  \)
and \( \tilde{m}_{l,m,l',m'}^{b,i,j}\left( \tilde{\beta },\tilde{\mu }\right) \equiv \left( 2m^{*}\right) ^{-1}m_{l,m,l',m'}^{b,i,j}\left( \beta ,\mu \right)  \)
and we have ommitted the dependence of some of the functions defined
above on \( \tilde{\beta },\tilde{\mu } \) for brevity. \bibliographystyle{apsrev}
\bibliography{bibliography}

\begin{thebibliography}{53}
\expandafter\ifx\csname natexlab\endcsname\relax\def\natexlab#1{#1}\fi
\expandafter\ifx\csname bibnamefont\endcsname\relax
  \def\bibnamefont#1{#1}\fi
\expandafter\ifx\csname bibfnamefont\endcsname\relax
  \def\bibfnamefont#1{#1}\fi
\expandafter\ifx\csname citenamefont\endcsname\relax
  \def\citenamefont#1{#1}\fi
\expandafter\ifx\csname url\endcsname\relax
  \def\url#1{\texttt{#1}}\fi
\expandafter\ifx\csname urlprefix\endcsname\relax\def\urlprefix{URL }\fi
\providecommand{\bibinfo}[2]{#2}
\providecommand{\eprint}[2][]{\url{#2}}

\bibitem[{\citenamefont{Bardeen et~al.}(1957)\citenamefont{Bardeen, Cooper, and
  Schrieffer}}]{Bardeen-Cooper-Schrieffer-57}
\bibinfo{author}{\bibfnamefont{J.}~\bibnamefont{Bardeen}},
  \bibinfo{author}{\bibfnamefont{L.}~\bibnamefont{Cooper}}, \bibnamefont{and}
  \bibinfo{author}{\bibfnamefont{J.}~\bibnamefont{Schrieffer}},
  \bibinfo{journal}{Phys. Rev.} \textbf{\bibinfo{volume}{108}},
  \bibinfo{pages}{1175} (\bibinfo{year}{1957}).

\bibitem[{\citenamefont{Balian}(1964)}]{Balian-64}
\bibinfo{author}{\bibfnamefont{R.}~\bibnamefont{Balian}}, in
  \emph{\bibinfo{booktitle}{The Many-Body Problem, Vol.2}}, edited by
  \bibinfo{editor}{\bibfnamefont{E.}~\bibnamefont{Caianiello}}
  (\bibinfo{year}{1964}).

\bibitem[{\citenamefont{Micnas et~al.}(1990)\citenamefont{Micnas, Ranninger,
  and Robaskiewicz}}]{Micnas-Ranninger-Robaskiewicz-90}
\bibinfo{author}{\bibfnamefont{R.}~\bibnamefont{Micnas}},
  \bibinfo{author}{\bibfnamefont{J.}~\bibnamefont{Ranninger}},
  \bibnamefont{and}
  \bibinfo{author}{\bibfnamefont{S.}~\bibnamefont{Robaskiewicz}},
  \bibinfo{journal}{Rev. Mod. Phys.} \textbf{\bibinfo{volume}{62}},
  \bibinfo{pages}{113} (\bibinfo{year}{1990}).

\bibitem[{\citenamefont{Landau and Lifshitz}(1958)}]{Landau-Lifshitz-QM}
\bibinfo{author}{\bibfnamefont{L.}~\bibnamefont{Landau}} \bibnamefont{and}
  \bibinfo{author}{\bibfnamefont{E.}~\bibnamefont{Lifshitz}},
  \emph{\bibinfo{title}{Quantum Mechanics, non-Relativistic Theory}}
  (\bibinfo{publisher}{Addison-Wesley}, \bibinfo{year}{1958}).

\bibitem[{\citenamefont{Eagles}(1969)}]{Eagles-69}
\bibinfo{author}{\bibfnamefont{D.}~\bibnamefont{Eagles}},
  \bibinfo{journal}{Phys. Rev.} \textbf{\bibinfo{volume}{186}},
  \bibinfo{pages}{456} (\bibinfo{year}{1969}).

\bibitem[{\citenamefont{Pincus et~al.}(1973)\citenamefont{Pincus, Chaikin, and
  C.F.~Coll}}]{Pincus-et-al-73}
\bibinfo{author}{\bibfnamefont{P.}~\bibnamefont{Pincus}},
  \bibinfo{author}{\bibfnamefont{P.}~\bibnamefont{Chaikin}}, \bibnamefont{and}
  \bibinfo{author}{\bibfnamefont{I.}~\bibnamefont{C.F.~Coll}},
  \bibinfo{journal}{Solid State Communications} \textbf{\bibinfo{volume}{12}},
  \bibinfo{pages}{1265} (\bibinfo{year}{1973}).

\bibitem[{\citenamefont{Leggett}(1980)}]{Leggett-80}
\bibinfo{author}{\bibfnamefont{A.}~\bibnamefont{Leggett}}, in
  \emph{\bibinfo{booktitle}{Modern trends in the theory of condensed matter}},
  edited by \bibinfo{editor}{\bibfnamefont{A.}~\bibnamefont{Pekelski}}
  \bibnamefont{and} \bibinfo{editor}{\bibfnamefont{J.}~\bibnamefont{Przystawa}}
  (\bibinfo{publisher}{Springer-Verlag, Berlin}, \bibinfo{year}{1980}).

\bibitem[{\citenamefont{Nozières and
  Schmitt-Rink}(1985)}]{Nozieres-SchmittRink-85}
\bibinfo{author}{\bibfnamefont{P.}~\bibnamefont{Nozières}} \bibnamefont{and}
  \bibinfo{author}{\bibfnamefont{S.}~\bibnamefont{Schmitt-Rink}},
  \bibinfo{journal}{J. Low Temp. Phys.} \textbf{\bibinfo{volume}{59}},
  \bibinfo{pages}{195} (\bibinfo{year}{1985}).

\bibitem[{\citenamefont{Engelbrecht et~al.}(1998)\citenamefont{Engelbrecht,
  Nazarenko, and Randeria}}]{Engelbrecht-Nazarenko-Randeria-98}
\bibinfo{author}{\bibfnamefont{J.~R.} \bibnamefont{Engelbrecht}},
  \bibinfo{author}{\bibfnamefont{A.}~\bibnamefont{Nazarenko}},
  \bibnamefont{and} \bibinfo{author}{\bibfnamefont{M.}~\bibnamefont{Randeria}},
  \bibinfo{journal}{Phys. Rev. B} \textbf{\bibinfo{volume}{57}},
  \bibinfo{pages}{13406} (\bibinfo{year}{1998}).

\bibitem[{\citenamefont{den Hertog}(1999)}]{DenHertog-99}
\bibinfo{author}{\bibfnamefont{B.~C.} \bibnamefont{den Hertog}},
  \bibinfo{journal}{Phys. Rev. B} \textbf{\bibinfo{volume}{60}},
  \bibinfo{pages}{559} (\bibinfo{year}{1999}).

\bibitem[{\citenamefont{Andrenacci et~al.}(1999)\citenamefont{Andrenacci,
  Perali, Pieri, and Strinati}}]{Andrenacci-Perali-Pieri-Strinati-99}
\bibinfo{author}{\bibfnamefont{N.}~\bibnamefont{Andrenacci}},
  \bibinfo{author}{\bibfnamefont{A.}~\bibnamefont{Perali}},
  \bibinfo{author}{\bibfnamefont{P.}~\bibnamefont{Pieri}}, \bibnamefont{and}
  \bibinfo{author}{\bibfnamefont{G.}~\bibnamefont{Strinati}},
  \bibinfo{journal}{Phys. Rev. B} \textbf{\bibinfo{volume}{60}},
  \bibinfo{pages}{12410} (\bibinfo{year}{1999}).

\bibitem[{\citenamefont{Wallington and Annett}(2000)}]{Jon-James-2000}
\bibinfo{author}{\bibfnamefont{J.~P.} \bibnamefont{Wallington}}
  \bibnamefont{and} \bibinfo{author}{\bibfnamefont{J.~F.}
  \bibnamefont{Annett}}, \bibinfo{journal}{Phys. Rev. B}
  \textbf{\bibinfo{volume}{61}}, \bibinfo{pages}{1433} (\bibinfo{year}{2000}).

\bibitem[{\citenamefont{Chen et~al.}(2000)\citenamefont{Chen, Kosztin, and
  Levin}}]{Chen-Kosztin-Levin-00}
\bibinfo{author}{\bibfnamefont{Q.}~\bibnamefont{Chen}},
  \bibinfo{author}{\bibfnamefont{I.}~\bibnamefont{Kosztin}}, \bibnamefont{and}
  \bibinfo{author}{\bibfnamefont{K.}~\bibnamefont{Levin}},
  \bibinfo{journal}{Phys. Rev. Lett.} \textbf{\bibinfo{volume}{85}},
  \bibinfo{pages}{2801} (\bibinfo{year}{2000}).

\bibitem[{\citenamefont{Soares et~al.}(2002)\citenamefont{Soares, Kokubun,
  Rodr{\'i}guez-N{\'u\~n}ez, and
  Rend{\'o}n}}]{Soares-Kokubun-RodriguezNunez-Rendon-02}
\bibinfo{author}{\bibfnamefont{M.~B.} \bibnamefont{Soares}},
  \bibinfo{author}{\bibfnamefont{F.}~\bibnamefont{Kokubun}},
  \bibinfo{author}{\bibfnamefont{J.~J.}
  \bibnamefont{Rodr{\'i}guez-N{\'u\~n}ez}}, \bibnamefont{and}
  \bibinfo{author}{\bibfnamefont{O.}~\bibnamefont{Rend{\'o}n}},
  \bibinfo{journal}{Phys. Rev. B} \textbf{\bibinfo{volume}{65}},
  \bibinfo{pages}{174506} (\bibinfo{year}{2002}).

\bibitem[{\citenamefont{Stintzing and Zwerger}(1997)}]{Zwerger-97}
\bibinfo{author}{\bibfnamefont{S.}~\bibnamefont{Stintzing}} \bibnamefont{and}
  \bibinfo{author}{\bibfnamefont{W.}~\bibnamefont{Zwerger}},
  \bibinfo{journal}{Phys. Rev. B} \textbf{\bibinfo{volume}{56}},
  \bibinfo{pages}{9004} (\bibinfo{year}{1997}).

\bibitem[{\citenamefont{Gottfried}(1966)}]{Gottfried-66}
\bibinfo{author}{\bibfnamefont{K.}~\bibnamefont{Gottfried}},
  \emph{\bibinfo{title}{Quantum Mechanics}}, vol.~\bibinfo{volume}{I}
  (\bibinfo{publisher}{W.A. Benjamin, New York}, \bibinfo{year}{1966}).

\bibitem[{\citenamefont{Villarroel}(1998)}]{Villarroel-98}
\bibinfo{author}{\bibfnamefont{D.}~\bibnamefont{Villarroel}},
  \bibinfo{journal}{Eur. J. Phys.} \textbf{\bibinfo{volume}{19}},
  \bibinfo{pages}{85} (\bibinfo{year}{1998}).

\bibitem[{\citenamefont{Quintanilla and
  Gyorffy}(2000)}]{Quintanilla-Gyorffy-00}
\bibinfo{author}{\bibfnamefont{J.}~\bibnamefont{Quintanilla}} \bibnamefont{and}
  \bibinfo{author}{\bibfnamefont{B.}~\bibnamefont{Gyorffy}},
  \bibinfo{journal}{Physica B} \textbf{\bibinfo{volume}{284-288}},
  \bibinfo{pages}{421} (\bibinfo{year}{2000}),
  \bibinfo{note}{cond-mat/9909052}.

\bibitem[{\citenamefont{Babev and Kleinert}(1998)}]{Babaev-Kleinert-98}
\bibinfo{author}{\bibfnamefont{E.}~\bibnamefont{Babev}} \bibnamefont{and}
  \bibinfo{author}{\bibfnamefont{H.}~\bibnamefont{Kleinert}},
  \bibinfo{journal}{Phys. Lett. B} \textbf{\bibinfo{volume}{438}},
  \bibinfo{pages}{311} (\bibinfo{year}{1998}).

\bibitem[{\citenamefont{Randeria et~al.}(1990)\citenamefont{Randeria, Duan, and
  Shieh}}]{Randeria-Duan-Shieh-90}
\bibinfo{author}{\bibfnamefont{M.}~\bibnamefont{Randeria}},
  \bibinfo{author}{\bibfnamefont{J.-M.} \bibnamefont{Duan}}, \bibnamefont{and}
  \bibinfo{author}{\bibfnamefont{L.-Y.} \bibnamefont{Shieh}},
  \bibinfo{journal}{Phys. Rev. B} \textbf{\bibinfo{volume}{41}},
  \bibinfo{pages}{327} (\bibinfo{year}{1990}).

\bibitem[{\citenamefont{Drechsler and Zwerger}(1992)}]{Zwerger-92}
\bibinfo{author}{\bibfnamefont{M.}~\bibnamefont{Drechsler}} \bibnamefont{and}
  \bibinfo{author}{\bibfnamefont{W.}~\bibnamefont{Zwerger}},
  \bibinfo{journal}{Ann. Phys.} \textbf{\bibinfo{volume}{1}},
  \bibinfo{pages}{15} (\bibinfo{year}{1992}).

\bibitem[{\citenamefont{de~Melo et~al.}(1993)\citenamefont{de~Melo, Randeria,
  and Engelbrecht}}]{SaDeMelo-Randeria-Engelbrecht-93}
\bibinfo{author}{\bibfnamefont{C.~S.} \bibnamefont{de~Melo}},
  \bibinfo{author}{\bibfnamefont{M.}~\bibnamefont{Randeria}}, \bibnamefont{and}
  \bibinfo{author}{\bibfnamefont{J.}~\bibnamefont{Engelbrecht}},
  \bibinfo{journal}{Phys. Rev. Lett.} \textbf{\bibinfo{volume}{71}},
  \bibinfo{pages}{3202} (\bibinfo{year}{1993}).

\bibitem[{\citenamefont{Engelbrecht et~al.}(1997)\citenamefont{Engelbrecht,
  Randeria, and de~Melo}}]{Engelbrecht-Randeria-SaDeMelo-97}
\bibinfo{author}{\bibfnamefont{J.}~\bibnamefont{Engelbrecht}},
  \bibinfo{author}{\bibfnamefont{M.}~\bibnamefont{Randeria}}, \bibnamefont{and}
  \bibinfo{author}{\bibfnamefont{C.~S.} \bibnamefont{de~Melo}},
  \bibinfo{journal}{Phys. Rev. B} \textbf{\bibinfo{volume}{55}},
  \bibinfo{pages}{15153} (\bibinfo{year}{1997}).

\bibitem[{\citenamefont{Marini et~al.}(1998)\citenamefont{Marini, Pistolesi,
  and Strinati}}]{Marini-Pistolesi-Strinati-98}
\bibinfo{author}{\bibfnamefont{M.}~\bibnamefont{Marini}},
  \bibinfo{author}{\bibfnamefont{F.}~\bibnamefont{Pistolesi}},
  \bibnamefont{and} \bibinfo{author}{\bibfnamefont{G.}~\bibnamefont{Strinati}},
  \bibinfo{journal}{Eur. Phys. J. B} \textbf{\bibinfo{volume}{1}},
  \bibinfo{pages}{151} (\bibinfo{year}{1998}).

\bibitem[{\citenamefont{Haussmann}(1993)}]{Haussmann-93}
\bibinfo{author}{\bibfnamefont{R.}~\bibnamefont{Haussmann}},
  \bibinfo{journal}{Z. Phys. B} \textbf{\bibinfo{volume}{91}},
  \bibinfo{pages}{291} (\bibinfo{year}{1993}).

\bibitem[{\citenamefont{Haussmann}(1994)}]{Haussmann-94}
\bibinfo{author}{\bibfnamefont{R.}~\bibnamefont{Haussmann}},
  \bibinfo{journal}{Phys. Rev. B} \textbf{\bibinfo{volume}{49}},
  \bibinfo{pages}{12975} (\bibinfo{year}{1994}).

\bibitem[{\citenamefont{Babev and Kleinert}(1999)}]{Babaev-Kleinert-99}
\bibinfo{author}{\bibfnamefont{E.}~\bibnamefont{Babev}} \bibnamefont{and}
  \bibinfo{author}{\bibfnamefont{H.}~\bibnamefont{Kleinert}},
  \bibinfo{journal}{Phys. Rev. B} \textbf{\bibinfo{volume}{59}},
  \bibinfo{pages}{12083} (\bibinfo{year}{1999}).

\bibitem[{\citenamefont{Pistolesi and Strinati}(1994)}]{Pistolesi-Strinati-94}
\bibinfo{author}{\bibfnamefont{F.}~\bibnamefont{Pistolesi}} \bibnamefont{and}
  \bibinfo{author}{\bibfnamefont{G.}~\bibnamefont{Strinati}},
  \bibinfo{journal}{Phys. Rev. B} \textbf{\bibinfo{volume}{49}},
  \bibinfo{pages}{6356} (\bibinfo{year}{1994}).

\bibitem[{\citenamefont{Pistolesi and Strinati}(1996)}]{Pistolesi-Strinati-96}
\bibinfo{author}{\bibfnamefont{F.}~\bibnamefont{Pistolesi}} \bibnamefont{and}
  \bibinfo{author}{\bibfnamefont{G.}~\bibnamefont{Strinati}},
  \bibinfo{journal}{Phys. Rev. B} \textbf{\bibinfo{volume}{53}},
  \bibinfo{pages}{15168} (\bibinfo{year}{1996}).

\bibitem[{\citenamefont{Randeria}(1995)}]{Randeria-95}
\bibinfo{author}{\bibfnamefont{M.}~\bibnamefont{Randeria}}, in
  \emph{\bibinfo{booktitle}{{Bose}-{Einstein} Condensation}}, edited by
  \bibinfo{editor}{\bibfnamefont{A.}~\bibnamefont{Griffin}},
  \bibinfo{editor}{\bibfnamefont{D.}~\bibnamefont{Snorke}}, \bibnamefont{and}
  \bibinfo{editor}{\bibfnamefont{S.}~\bibnamefont{Stringari}}
  (\bibinfo{publisher}{Cambridge University Press}, \bibinfo{year}{1995}).

\bibitem[{\citenamefont{Alexandrov and Rubin}(1993)}]{Alexandrov-Rubin-93}
\bibinfo{author}{\bibfnamefont{A.}~\bibnamefont{Alexandrov}} \bibnamefont{and}
  \bibinfo{author}{\bibfnamefont{S.}~\bibnamefont{Rubin}},
  \bibinfo{journal}{Phys. Rev. B} \textbf{\bibinfo{volume}{47}},
  \bibinfo{pages}{5141} (\bibinfo{year}{1993}).

\bibitem[{\citenamefont{Quintanilla}(2001)}]{Quintanilla-01}
\bibinfo{author}{\bibfnamefont{J.}~\bibnamefont{Quintanilla}}, Ph.D. thesis,
  \bibinfo{school}{University of Bristol} (\bibinfo{year}{2001}).

\bibitem[{\citenamefont{Annett et~al.}(1996)\citenamefont{Annett, Goldenfeld,
  and Leggett}}]{Annett-Goldenfeld-Leggett-96}
\bibinfo{author}{\bibfnamefont{J.~F.} \bibnamefont{Annett}},
  \bibinfo{author}{\bibfnamefont{N.}~\bibnamefont{Goldenfeld}},
  \bibnamefont{and} \bibinfo{author}{\bibfnamefont{A.}~\bibnamefont{Leggett}},
  in \emph{\bibinfo{booktitle}{Physical properties of high-temperature
  superconductors}}, edited by
  \bibinfo{editor}{\bibfnamefont{D.}~\bibnamefont{Ginsberg}}
  (\bibinfo{publisher}{World Scientific}, \bibinfo{year}{1996}),
  vol.~\bibinfo{volume}{V}.

\bibitem[{\citenamefont{Anderson and Brinkman}(1975)}]{Anderson-Brinkman-74}
\bibinfo{author}{\bibfnamefont{P.}~\bibnamefont{Anderson}} \bibnamefont{and}
  \bibinfo{author}{\bibfnamefont{W.}~\bibnamefont{Brinkman}}, in
  \emph{\bibinfo{booktitle}{The Helium Liquids: Proceedings of the 15th
  Scottish Universities Summer School in Physics, 1974}}, edited by
  \bibinfo{editor}{\bibfnamefont{J.}~\bibnamefont{Armitage}} \bibnamefont{and}
  \bibinfo{editor}{\bibfnamefont{I.}~\bibnamefont{Farquhar}}
  (\bibinfo{publisher}{Academic Press, New York}, \bibinfo{year}{1975}),
  \bibinfo{note}{[Reprinted in P.W. Anderson, {"Basic Notions of Condensed
  Matter Physics"} (Benjamin/Cummings, 1984)]}.

\bibitem[{\citenamefont{Nagaosa}(1999)}]{Nagaosa-99}
\bibinfo{author}{\bibfnamefont{N.}~\bibnamefont{Nagaosa}},
  \emph{\bibinfo{title}{Quantum Field Theory in Condensed Matter Physics}}
  (\bibinfo{publisher}{Springer-Verlag}, \bibinfo{year}{1999}).

\bibitem[{\citenamefont{Negele and Orland}(1988)}]{Negele-Orland-88}
\bibinfo{author}{\bibfnamefont{J.}~\bibnamefont{Negele}} \bibnamefont{and}
  \bibinfo{author}{\bibfnamefont{H.}~\bibnamefont{Orland}},
  \emph{\bibinfo{title}{Quantum Many-Particle Systems}}
  (\bibinfo{publisher}{Addison-Wesley}, \bibinfo{year}{1988}).

\bibitem[{\citenamefont{Hubbard}(1959)}]{Hubbard-59}
\bibinfo{author}{\bibfnamefont{J.}~\bibnamefont{Hubbard}},
  \bibinfo{journal}{Phys. Rev. Lett.} \textbf{\bibinfo{volume}{3}},
  \bibinfo{pages}{77} (\bibinfo{year}{1959}).

\bibitem[{\citenamefont{Pieri and Strinati}(2000)}]{Pieri-Strinati-00}
\bibinfo{author}{\bibfnamefont{P.}~\bibnamefont{Pieri}} \bibnamefont{and}
  \bibinfo{author}{\bibfnamefont{G.}~\bibnamefont{Strinati}},
  \bibinfo{journal}{Phys. Rev. B} \textbf{\bibinfo{volume}{61}},
  \bibinfo{pages}{15370} (\bibinfo{year}{2000}).

\bibitem[{\citenamefont{Gyorffy et~al.}(1991)\citenamefont{Gyorffy, Staunton,
  and Stocks}}]{Gyorffy-Staunton-Stocks-91}
\bibinfo{author}{\bibfnamefont{B.}~\bibnamefont{Gyorffy}},
  \bibinfo{author}{\bibfnamefont{J.}~\bibnamefont{Staunton}}, \bibnamefont{and}
  \bibinfo{author}{\bibfnamefont{G.}~\bibnamefont{Stocks}},
  \bibinfo{journal}{Phys. Rev. B} \textbf{\bibinfo{volume}{44}},
  \bibinfo{pages}{5190} (\bibinfo{year}{1991}).

\bibitem[{\citenamefont{Bak and Micnas}(2000)}]{Bak-Micnas-00}
\bibinfo{author}{\bibfnamefont{M.}~\bibnamefont{Bak}} \bibnamefont{and}
  \bibinfo{author}{\bibfnamefont{R.}~\bibnamefont{Micnas}},
  \bibinfo{journal}{Acta Physica Polonica A} \textbf{\bibinfo{volume}{97}}
  (\bibinfo{year}{2000}), \bibinfo{note}{proceedings of the European Conference
  Physics of Magnetism 99, Pozna{\'n} 1999}.

\bibitem[{\citenamefont{Quintanilla and
  Gyorffy}(2002)}]{Quintanilla-Gyorffy-02}
\bibinfo{author}{\bibfnamefont{J.}~\bibnamefont{Quintanilla}} \bibnamefont{and}
  \bibinfo{author}{\bibfnamefont{B.}~\bibnamefont{Gyorffy}},
  \bibinfo{journal}{J. Phys.: Condens. Matter} \textbf{\bibinfo{volume}{14}},
  \bibinfo{pages}{6591} (\bibinfo{year}{2002}),
  \bibinfo{note}{cond-mat/0106251}.

\bibitem[{\citenamefont{Garrido et~al.}(2001)\citenamefont{Garrido, Sarriguren,
  de~Guerra, Lombardo, Schuck, and
  Schulze}}]{Sarriguren-MoyaDeGuerra-Lombardo-Schuck-Schulze-01}
\bibinfo{author}{\bibfnamefont{E.}~\bibnamefont{Garrido}},
  \bibinfo{author}{\bibfnamefont{P.}~\bibnamefont{Sarriguren}},
  \bibinfo{author}{\bibfnamefont{E.~M.} \bibnamefont{de~Guerra}},
  \bibinfo{author}{\bibfnamefont{U.}~\bibnamefont{Lombardo}},
  \bibinfo{author}{\bibfnamefont{P.}~\bibnamefont{Schuck}}, \bibnamefont{and}
  \bibinfo{author}{\bibfnamefont{H.~J.} \bibnamefont{Schulze}},
  \bibinfo{journal}{Phys. Rev. C} \textbf{\bibinfo{volume}{63}},
  \bibinfo{pages}{037304} (\bibinfo{year}{2001}).

\bibitem[{\citenamefont{Bruun et~al.}(1999)\citenamefont{Bruun, Castin, Dum,
  and Burnett}}]{Bruun-Castin-Dum-Burnett-99}
\bibinfo{author}{\bibfnamefont{G.}~\bibnamefont{Bruun}},
  \bibinfo{author}{\bibfnamefont{Y.}~\bibnamefont{Castin}},
  \bibinfo{author}{\bibfnamefont{R.}~\bibnamefont{Dum}}, \bibnamefont{and}
  \bibinfo{author}{\bibfnamefont{K.}~\bibnamefont{Burnett}},
  \bibinfo{journal}{Eur. Phys. J. D} \textbf{\bibinfo{volume}{7}},
  \bibinfo{pages}{433} (\bibinfo{year}{1999}).

\bibitem[{\citenamefont{Combescot}(2001)}]{Combescot-01}
\bibinfo{author}{\bibfnamefont{R.}~\bibnamefont{Combescot}},
  \bibinfo{journal}{Europhys. Lett.} pp. \bibinfo{pages}{150--156}
  (\bibinfo{year}{2001}).

\bibitem[{\citenamefont{Holland et~al.}(2001)\citenamefont{Holland, Kokkelmans,
  Chiofalo, and Walser}}]{Holland-Kokkelmans-Chiofalo-Walser-01}
\bibinfo{author}{\bibfnamefont{M.}~\bibnamefont{Holland}},
  \bibinfo{author}{\bibfnamefont{S.~J. J. M.~F.} \bibnamefont{Kokkelmans}},
  \bibinfo{author}{\bibfnamefont{M.~L.} \bibnamefont{Chiofalo}},
  \bibnamefont{and} \bibinfo{author}{\bibfnamefont{R.}~\bibnamefont{Walser}},
  \bibinfo{journal}{Phys. Rev. Lett.} \textbf{\bibinfo{volume}{87}},
  \bibinfo{pages}{120406} (\bibinfo{year}{2001}).

\bibitem[{\citenamefont{Ohashi and Griffin}()}]{Ohashi-Griffin-02}
\bibinfo{author}{\bibfnamefont{Y.}~\bibnamefont{Ohashi}} \bibnamefont{and}
  \bibinfo{author}{\bibfnamefont{A.}~\bibnamefont{Griffin}},
  \bibinfo{note}{cond-mat/0201262}.

\bibitem[{\citenamefont{Jin}(2002)}]{Jin-02}
\bibinfo{author}{\bibfnamefont{D.}~\bibnamefont{Jin}},
  \bibinfo{journal}{Physics World} \textbf{\bibinfo{volume}{15}},
  \bibinfo{pages}{27} (\bibinfo{year}{2002}).

\bibitem[{\citenamefont{Lin and Hirsch}(1986)}]{Lin-Hirsch-86}
\bibinfo{author}{\bibfnamefont{H.}~\bibnamefont{Lin}} \bibnamefont{and}
  \bibinfo{author}{\bibfnamefont{J.}~\bibnamefont{Hirsch}},
  \bibinfo{journal}{Phys. Rev. B} \textbf{\bibinfo{volume}{33}},
  \bibinfo{pages}{8155} (\bibinfo{year}{1986}).

\bibitem[{\citenamefont{R{\"o}pke et~al.}(1998)\citenamefont{R{\"o}pke,
  Schnell, Schuck, and Nozi{\`e}res}}]{Ropke-Schnell-Schuck-Nozieres-98}
\bibinfo{author}{\bibfnamefont{G.}~\bibnamefont{R{\"o}pke}},
  \bibinfo{author}{\bibfnamefont{A.}~\bibnamefont{Schnell}},
  \bibinfo{author}{\bibfnamefont{P.}~\bibnamefont{Schuck}}, \bibnamefont{and}
  \bibinfo{author}{\bibfnamefont{P.}~\bibnamefont{Nozi{\`e}res}},
  \bibinfo{journal}{Phys. Rev. Lett.} \textbf{\bibinfo{volume}{80}},
  \bibinfo{pages}{3177} (\bibinfo{year}{1998}).

\bibitem[{\citenamefont{Burgy et~al.}(2001)\citenamefont{Burgy, Mayr,
  Martin-Mayor, Moreo, and Dagotto}}]{Burgy-Mayr-MartinMayor-Moreo-Dagotto-01}
\bibinfo{author}{\bibfnamefont{J.}~\bibnamefont{Burgy}},
  \bibinfo{author}{\bibfnamefont{M.}~\bibnamefont{Mayr}},
  \bibinfo{author}{\bibfnamefont{V.}~\bibnamefont{Martin-Mayor}},
  \bibinfo{author}{\bibfnamefont{A.}~\bibnamefont{Moreo}}, \bibnamefont{and}
  \bibinfo{author}{\bibfnamefont{E.}~\bibnamefont{Dagotto}},
  \bibinfo{journal}{Phys. Rev. Lett.} \textbf{\bibinfo{volume}{87}},
  \bibinfo{pages}{277202} (\bibinfo{year}{2001}).

\bibitem[{\citenamefont{Strange}(1998)}]{Strange-98}
\bibinfo{author}{\bibfnamefont{P.}~\bibnamefont{Strange}},
  \emph{\bibinfo{title}{Relativistic Quantum Mechanics}}
  (\bibinfo{publisher}{Cambridge University Press}, \bibinfo{year}{1998}).

\bibitem[{\citenamefont{Kohn and Luttinger}(1965)}]{Kohn-Luttinger-65}
\bibinfo{author}{\bibfnamefont{W.}~\bibnamefont{Kohn}} \bibnamefont{and}
  \bibinfo{author}{\bibfnamefont{J.}~\bibnamefont{Luttinger}},
  \bibinfo{journal}{Phys. Rev. Lett.} \textbf{\bibinfo{volume}{15}},
  \bibinfo{pages}{524} (\bibinfo{year}{1965}).

\bibitem[{\citenamefont{Chubukov and Kagan}(1989)}]{Chubukov-Kagan-89}
\bibinfo{author}{\bibfnamefont{A.~V.} \bibnamefont{Chubukov}} \bibnamefont{and}
  \bibinfo{author}{\bibfnamefont{M.~Y.} \bibnamefont{Kagan}},
  \bibinfo{journal}{J. Phys.: Condens. Matter} \textbf{\bibinfo{volume}{1}},
  \bibinfo{pages}{3135} (\bibinfo{year}{1989}).

\end{thebibliography}

\end{document}